\documentclass[draftclsnofoot, onecolumn, 12pt]{IEEEtran}

\IEEEoverridecommandlockouts

\usepackage{cite}
\usepackage{graphicx}
\usepackage{psfrag} 
\usepackage{bm}
\usepackage{subfigure}
\usepackage{amsmath,amssymb,amsfonts}
\usepackage{epsfig}
\usepackage{amsmath}
\usepackage{multirow}
\usepackage{lineno}
\usepackage{algorithm}
\usepackage{algorithmic} 
\usepackage{diagbox} 
\usepackage[margin=0.7in]{geometry}
\usepackage{lipsum}

\usepackage{booktabs}

\usepackage{xcolor}

\newcommand{\p}[1]{\mathop{\mbox{\it p} } }

\renewcommand{\vec}[1]{\ensuremath{\boldsymbol{#1}}}
\newcommand{\be}{\begin{equation}}
	\newcommand{\ee}{\end{equation}}
\newcommand{\ba}{\begin{array}}
	\newcommand{\ea}{\end{array}}
\newcommand{\bea}{\begin{eqnarray}}
	\newcommand{\eea}{\end{eqnarray}}
\newcommand{\bean}{\begin{eqnarray*}}
	\newcommand{\eean}{\end{eqnarray*}}

\newcommand{\rmh}{^{\rm \dag}}

\renewcommand{\Re}{\mathop{\rm Re}}
\renewcommand{\Im}{\mathop{\rm Im}}

\definecolor{white}{rgb}{1,1,1}

\newtheorem{property}{\bf Property}

\title{Optimal Power Allocation and AI Receiver Design for Superimposed DMRS and Data Transmission}
\author{Sha Hu and Zhongwang Fu\\
\thanks{The authors work with Nyquist Research Center, Huawei Technologies Sweden AB, Sweden, and Zhongwang Fu was an intern during the time of this work~\cite{fu26}. Email: \{hu.sha,\;zhongwang.fu\}@huawei.com.} 
}

\begin{document}
\maketitle

\begin{abstract}

In this paper, we consider transmissions with superimposed (SI) demodulation-reference-symbol (DMRS) and data in orthogonal frequency-division multiplexing (OFDM) based multiple-input multiple-output (MIMO) systems. First, we derive an analytical framework to characterize the iterative behaviour between the mean-square errors (MSEs) of channel estimation (CE) and MIMO detection (MD) within an iterative CE and detection (ICED) process. This framework is subsequently utilized to optimize power allocation and pilot patterns between the DMRS and data symbols for SI-DMRS transmission. Second, we design an artificial intelligence (AI) based receiver built upon Transformer encoders for SI-DMRS transmissions, which incorporates an iterative CE and detection (ICED) structure. Simulation results demonstrate that the proposed AI-ICED receiver, combined with SI-DMRS, effectively increases spectral efficiency (SE) compared to conventional systems using non-overlapped DMRS and data symbols.

\end{abstract}

\section{Introduction}

Driven by evolutions towards the sixth-generation radio (6GR) networks, artificial intelligence (AI) and deep neural-network (NN) driven paradigm shifts have became a broad consensus~\cite{Wang2026AI, Wu2024, Sajid2026, lin2023, OH17, QJ19}. Future wireless networks are steadily transitioning towards  AI-native designs, including AI transceiver designs for multiple-input multiple-output (MIMO) and orthogonal frequency-division multiplexing (OFDM) systems. Unlike a conventional receiver design that relies on a cascade of independently optimized modules, AI-based receiver can unify multiple core functionalities such as channel estimation (CE) and MIMO detection (MD)~\cite{Li2025, QS24, QS26, Hu26, BB26, HW98, C23, HL18, MA24} to harvest joint processing gains.

Although classical algorithms such as linear minimum mean square error (LMMSE) estimator and quasi-maximum likelihood detector (MLD)~\cite{DD04, HR17} demonstrate good performance, exploring AI design is still of interest. Firstly, the inference process naturally aligns with contemporary parallel computing hardware, offering a path to reduce computational latency. Secondly, a unified NN architecture can mitigate information barriers among different modules, achieving joint processing gains through latent feature exchanges, thereby demonstrating a potential to surpass conventional baselines in challenge scenarios.

A direction that AI could be beneficial in 6GR system is to optimize the overheads of pilots to increase spectral efficiency (SE). A classical dilemma is that to attain a good CE accuracy, more pilots should be transmitted but then the SE is decreased, and vice-visa, with fewer pilots the CE is not good enough and may degrade the data detection performance. In current fifth-generation new-radio (5G-NR) system, demodulation reference signal (DMRS) as pilots are transmitted on stand-alone resource elements (RE) to avoid interfering data transmissions. To increase SE, superimposed DMRS (SI-DMRS) has been proposed~\cite{SB25, UV17, LJ25} which directly transmits DMRS overlapped with data symbols. This yields zero overheads from pilots, but the challenge is to attain an accurate CE from SI-DMRS and data\footnote{Note that in our framework introduced later, the current DMRS patterns in 5G-NR can be seen as special cases of SI-DMRS by setting the power allocation of data symbols to zero on REs with SI-DMRS.}. While reducing pilot overheads or adopting superimposed DMRS (SI-DMRS) directly increases SE, it degrades channel state information (CSI) accuracy, which compromises symbol detection. Although attention-based NN architectures have been proposed to enhance CE under limited pilots~\cite{QS26}, questions of how to optimize the balance between DMRS overheads and data detection, and design an AI receiver that is capable of reliable data recovery with a minimal pilot overheads, remain inadequately addressed.

Although conventional iterative CE and detection (ICED)~\cite{ying26, AH08} can be applied, the two tasks of CE and MD on REs with SI-DMRS are entangled with each other, and it is challenging to achieve satisfying performances for both. However, this type of problem is perfectly for AI to attack, which learns from training dataset to approach the performance bound of joint CE and MD that is infeasible with conventional methods. On the other hand, with the rapid iteration of large foundation models centred on Transformers (e.g., Qwen, DeepSeek, Kimi~\cite{Qwen25, ds25, kim2.5}), in-context learning (ICL) has attracted widespread attention~\cite{SS25, ZS24, SR26}. ICL essentially provides an unprecedented few-shot adaptation paradigm. Mapping this to communications systems, a few pilot signals and their RE-locations can serve as contextual information for interpreting the channel state information (CSI) and data-detecting of the received MIMO-OFDM grids. 

Motivated by challenges with SI-DMRS and inspirations from ICL advances, we propose an AI-ICED receiver architecture for resolving CE and MD with SI-DMRS transmissions in MIMO-OFDM systems, which leverages advanced Transformer encoders to provide a unified design for diverse transmission schemes that achieves near-optimal performance. The architecture adopts an unfolded iterative cascade with a Transformer-encoder based backbone that incorporates Virtual Width Networks (VWN)~\cite{Seed25}, Tensor Product Attention (TPA)~\cite{YZ26}, and Mixture-of-Experts (MoE)~\cite{ds24} layers. Between iteration stages, a differentiable physics-guided feature construction (PGFC) module converts intermediate predictions into explicit physical features and are forwarded to subsequent stages. 

The main contributions of this paper are summarized as follows:
\begin{itemize}
    \item \textbf{Theoretical analysis of power and DMRS allocations:} We have derived the MSEs for CE and MD within the ICED structure, proving the existence of an equilibrium state across iterations. Consequently, determining optimal power and DMRS resource allocations for general settings becomes straightforward. Furthermore, we incorporated mutual information effective signal-to-noise ratio mapping (MIESM)~\cite{YZ14} to effectively combine the two RE subsets for data detections, with and without superimposed DMRS, to evaluate final decoding performance, thereby enhancing the practical applicability of the proposed optimizations.
    
    \item \textbf{Transformer unified AI-ICED receiver design:} We have designed an AI-ICED receiver using Transformer encoder blocks as its core backbone. The AI receiver processes the received signal, DMRS symbols, and a pilot mask to output symbol likelihoods at each stage. In the final stage, bit log-likelihood ratios (LLRs) are computed and passed to the decoder to evaluate block error rate (BLER) performance. The key features of the proposed AI-ICED receiver are twofold: joint CE and MD at each iteration stage, and a physics-guided feature construction (PGFC) mechanism between iterations that enables progressive and joint refinements.

    \item \textbf{Thorough simulation results and benchmarks:} We have evaluated the proposed AI-ICED receiver against optimal MLD and genie-CSI-aided benchmarks. The results quantify the effectiveness of AI receiver under superimposed DMRS (SI-DMRS) schemes, demonstrating near-optimal MD performance with genie-CSI. Furthermore, the proposed receiver converges rapidly within 2 to 3 iterations and achieves substantial SE gains by mitigating DMRS overhead compared to a standard 5G-NR baseline.
\end{itemize}

The remainder of this paper is organized as follows. Sec.~II introduces the MIMO-OFDM system model with SI-DMRS, and a theoretical framework for analysing the MSEs of CE and MD in conventional ICED process is developed. Sec.~III investigated the optimal power allocation of SI-DMRS based on the analytical framework established in Sec.~II. Sec.~IV presents the proposed AI-ICED receiver design in detail. Sec.~V provides the simulation results, and Sec~VI concludes the paper.

\subsubsection*{Notations:} The Hermitian of a matrix $\vec{A}$ is denoted as $\vec{A}\rmh$, and the identity matrix is denoted as $\vec{I}$. The expectation operator is $\mathbb{E}\{\cdot\}$. The variables $B$, $N_{\mathrm{r}}$, $N_{\mathrm{t}}$, $N_{\mathrm{sym}}$, and $N_{\mathrm{sc}}$ denote the batch size, number of receive antennas, number of transmit antennas, number of OFDM symbols within one subframe that carries data (excludes OFDM symbols carrying control channel), and number of subcarriers for the considered bandwidth, respectively. The variables $N$ and $M$ denotes the number of REs that only carries data symbols, and the number of REs carrying both SI-DMRS and data, respectively. Furthermore, $L\!=\!N_{\mathrm{sym}}N_{\mathrm{sc}}\!=\!N\!+\!M$ denote the input sequence length for AI receiver for a CE and MD occurrence. The power allcoation factor between SI-DMRS and data is denoted as $b$, and the power factor for data without SI-DMRS is set to $a$, both $0\!<\!a, b\!\leq\!1$. Furthermore, $d_{\mathrm{model}}$ is model-size in Transformer encoder blocks, and $d_{\mathrm{ff}}$ is (feed-forward network) FFN dimension.

\begin{figure}[t]
    \centering
    \hspace{0mm}
    \includegraphics[width=0.41\textwidth]{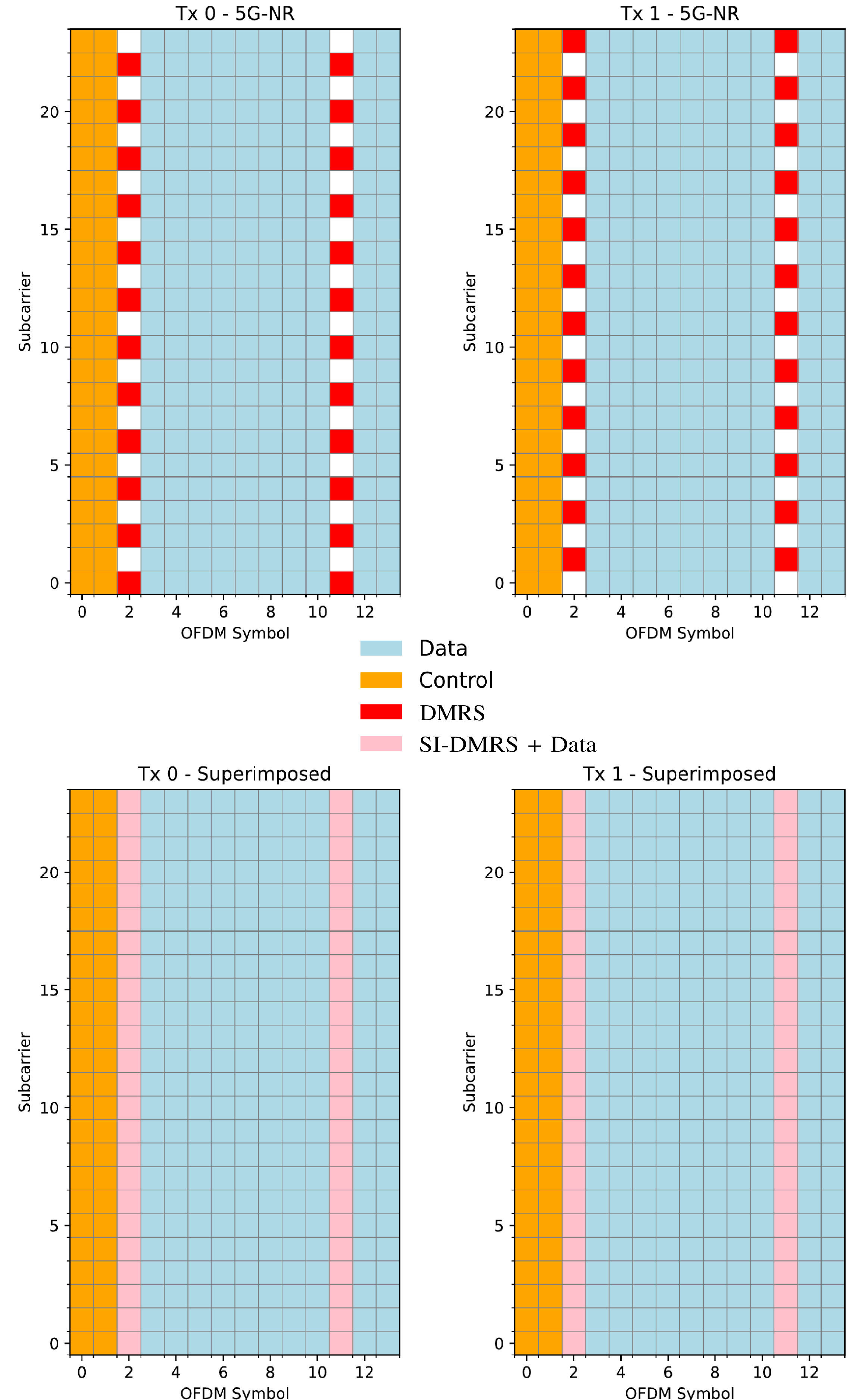}
           \vspace{-4mm}
    \caption{Example patterns of DMRS and data transmission in standard 5G-NR (top) versus with SI-DMRS (bottom). The bandwidth is 2 physical resource block (2PRB) with $N_{\text{sc}}\!=\!24$, and two OFDM symbols are used for control channel such that $N_{\text{sym}}\!=\!12$ for the SI-DMRS case.  In both cases, the transmit power per RE is normalized to unity. Under the SI-DMRS scheme, superimposing additional DMRS symbols onto data introduces no pilot overhead, however, because total transmit power is split between the DMRS and data components, both CE and MD become more challenging }
    \label{fig:pilot_strategies}
        \vspace{-6mm}
\end{figure}

\section{System Model and ICED Process}

\subsection{Received Signal Model}

We consider a MIMO-OFDM system consisting of $N_{\mathrm{t}}$ transmit-antennas (Tx) and $N_{\mathrm{r}}$ receiving-antennas (Rx). The transmitted signals are structured into a two-dimensional (2D) time-frequency resource grid, where transmission time interval (TTI) comprises $N_{\mathrm{sym}}$ OFDM symbols in the time domain and $N_{\mathrm{sc}}$ subcarriers in the frequency domain~\cite{3gpp38211}. Let $\vec{x}_{n,k}$ denote the transmit symbol vector mapped to the $n$th OFDM symbol and the $k$th subcarrier, with its entries drawn from an $M$-ary quadrature amplitude modulation ($M$-QAM) constellation. 

The corresponding received signal vector $\vec{y}_{n,k} \!\in\! \mathbb{C}^{N_{\mathrm{r}} \times 1}$ without DMRS is
\bea
    \vec{y}_{n,k} =\sqrt{\frac{1}{N_{\text{t}}}} \vec{H}_{n,k} \vec{x}_{n,k} + \vec{w}_{n,k}, 
\eea
where $\vec{H}_{n,k} \!\in\! \mathbb{C}^{N_{\mathrm{r}} \!\times\! N_{\mathrm{t}}}$ is the complex-valued frequency-domain MIMO channel matrix, and $\vec{w}_{n,k} \!\in\! \mathbb{C}^{N_{\mathrm{r}}\! \times\! 1}$ represents the additive white Gaussian noise (AWGN) at the receiver. The entries of $\vec{w}_{n,k}$ are independent and identically distributed (i.i.d.) complex Gaussian random variables distributed as $\vec{w}_{n,k} \!\sim\! \mathcal{CN}(\vec{0}, \sigma^2 \vec{I})$, where $\sigma^2$ denotes noise denotes the noise variance per receive antenna. For simplicity, we assume no spatial correlation, meaning the entries of $\vec{H}_{n,k}$ are also i.i.d. complex Gaussian random variables satisfying $\mathbb{E}\{\vec{H}_{n,k}\vec{H}_{n,k}\rmh\}\!=\!N_{\mathrm{t}}\vec{I}$. Furthermore, the data vectors are transmitted with unit-power such that $\mathbb{E}\{\vec{x}_{n,k}\vec{x}_{n,k}\rmh\}\!=\!\vec{I}$, and the received signal-to-noise ratio (SNR) is defined as $\mathrm{SNR} \!=\! 1 / \sigma^2$. By stacking all REs within one subframe, the received signal tensor $\vec{Y} \!\in\! \mathbb{C}^{N_{\mathrm{r}} \!\times\! N_{\mathrm{sym}} \!\times\! N_{\mathrm{sc}}}$ and the corresponding channel tensor $\vec{H} \!\in\! \mathbb{C}^{N_{\mathrm{r}} \!\times\! N_{\mathrm{t}} \!\times\! N_{\mathrm{sym}}\! \times\! N_{\mathrm{sc}}}$ constitute the complete observation space for the subsequent processing tasks.

To establish a framework for the iterative behaviours in ICED, we define some parameters as follows. Within one subframe, $N$ REs are allocated for data transmission without DMRS, and $M$ REs are allocated for SI-DMRS and data transmission. The total transmit power within one subframe is normalized to $N \!+\! M$. Further, denote $\mu \!=\! \frac{N}{M}$ and $k \!= \!1\! +\! \mu(1-a)$. 

The power allocation is structured as follows:
\begin{itemize}
    \item Data symbol vectors transmitted on REs without DMRS are with a power factor $a$.
    \item SI-DMRS symbols are transmitted with power $bk$.
    \item Data symbol vectors with SI-DMRS are transmitted with power $(1\!-\!b)k$.
\end{itemize}
Accordingly, the total power is fixed since $aN \!+\! kM \!= N \!+\! M$. The purpose of introduce the parameter $a$ is to analyse the potential benefits of borrowing powers from data symbols to boost the $SNR$ on REs with SI-DMRS, thereby the overall detection performance can be improved.

\subsection{CE with SI-DMRS}

On REs containing SI-DMRS, after removing the estimated data component, the received signal model for estimating the channel vector $\vec{h}$ for each Tx reads
\bea \label{ce-model}
\vec{y}_{\text{1}} =\sqrt{bk}\vec{h}s_{\text{DMRS}}+\sqrt{\frac{(1-b)k}{N_{\text{t}}}}(\vec{H}\vec{x}-\tilde{\vec{H}}\tilde{\vec{x}})+\vec{w},
\eea
where $\tilde{\vec{H}}$ and $\tilde{\vec{x}}$ are the estimated MIMO channel and data symbols, respectively, and $s$ is the transmitted DMRS symbol for a given Tx\footnote{We consider the case where DMRS for different Tx are non-overlapped and transmitted on separate REs. We also evaluated direct DMRS superimposition across Tx on the same RE (without orthogonal cover codes (OCC) as in 5G-NR), but found no noticeable performance gains, as it introduces spatial interference and reduces the effective DMRS power per Tx.}.  

Denote the CE error matrix as $\vec{\Delta\! H}\!=\! \vec{H}\!-\!\tilde{\vec{H}}$, with $\vec{\Delta h}$ denoting the column corresponding to a single transmit antenna. Likewise, define the detection error vector as $\vec{\Delta \!x}\!=\! \vec{x}\!-\!\tilde{\vec{x}}$. Assuming the CE-MSE and MD-MSE from the previous iteration are respectively defined as
\bea 
\mathbb{E}\{\vec{\Delta\! h} (\vec{\Delta \!h})\rmh \}\!&\!\!=\!\!&\! p\vec{I},\;\; 0\geq p\leq 1, \\
\mathbb{E}\{\vec{\Delta\! x} (\vec{\Delta \!x})\rmh \}\!&\!\!=\!\!&\! q\vec{I}, \;\; 0\geq d<\leq 1.
\eea 

In the next CE step in an ICED process, the update CE-MSE based on (\ref{ce-model}) can be computed as (see Appendix-A)
\bea \label{ce-mse}
\mathbb{E}\{\vec{\Delta\! h} (\vec{\Delta\! h})\rmh \}=\frac{k(1-b)(p+q-pq)+\sigma^2}{kb+k(1-b)(p+q-pq)+\sigma^2}\vec{I}.
\eea

\subsection{CE De-noising Among Pilots}

Note that the CE-MSE in (\ref{ce-mse}) is evaluated for a single RE, and typically an LMMSE filter is subsequently applied to de-noise the to  the CE over multiple REs across multiple REs carrying DMRS. In this case, the refined channel estimate is given by
\bea
\vec{h}_{\text{LMMSE}} = \vec{R}_{\text{h}}(\vec{R}_{\text{h}}+\tilde{\sigma}^2\vec{I})^{-1}\bar{\vec{h}},
\eea
where $\vec{R}_{\text{h}}$is the time-frequency channel correlation matrix for each Tx–Rx pair, the vector $\bar{\vec{h}}$ comprises the CE on $M/N_{\text{t}}$ REs for each link, and $\tilde{\sigma}^2$ is the effective noise power that incorporates both the noise and the residual data interference components. This formulation enables a maximum coherent processing gain of $M/N_{\text{t}}$. 

Although the practical gain can be analysed depends on $\vec{R}_{\text{h}}$, for simplicity we assume that $\mathbb{E}\{\vec{\Delta\! h} (\vec{\Delta \!h})\rmh \}$ is decreased by a factor $\frac{\gamma N_{\text{t}}}{M}$, where $\gamma\!\geq\!1$ accounts for any loss relative to the maximum coherent gain. Consequently, after de-noising, the CE-MSE is modelled as $p\vec{I}$, where
\bea  \label{p}
p=\bigg(\frac{\gamma N_{\text{t}}}{M}\bigg)\!\frac{k(1-b)(p+q-pq)+\sigma^2}{kb+k(1-b)(p+q-pq)+\sigma^2}.
\eea

\subsection{MD with SI-DMRS}

After CE is obtained and de-noised, after removing the DMRS component, the received signal model for detecting $\vec{x}$ is
\bea \label{md-model}
\vec{y}_{\text{2}} \!&\!\!\!\!=\!\!\!\!&\!\!\sqrt{bk}\vec{\Delta\! h}s_{\text{DMRS}}\!+\!\sqrt{\frac{(1\!-\!b)k}{N_{\text{t}}}}(\tilde{\vec{H}}\!+\!\vec{\Delta \!H})\vec{x}\!+\!\vec{w} \notag \\
&\!\!\!\!=\!\!\!\!&\!\!\sqrt{\frac{(1\!-\!b)k}{N_{\text{t}}}}\tilde{\vec{H}}\vec{x}\!+\!\sqrt{\frac{(1\!-\!b)k}{N_{\text{t}}}}\vec{\Delta\! H}\vec{x}\!+\!\sqrt{bk}\vec{\Delta \!h}s_{\text{DMRS}}\!+\!\vec{w}\!. \notag \\
\eea
For analytical tractability, the MD-MSE $\mathbb{E}\{\vec{\Delta x} (\vec{\Delta x})\rmh \}$ based on (\ref{md-model}) can be approximated as $q\vec{I}$ (see Appendix-B), where
\bea \label{q}
q=\frac{kp+\sigma^2}{k(1-b+pb)+\sigma^2}.
\eea

As seen, equations (\ref{p}) and (\ref{q}) define the equilibrium state of CE-MSE and MD-MSE. This is formalized in Property~1, with its proof provided in Appendix C.
\begin{property}
There exists a stable equilibrium state for CE-MSE and MD-MSE within the ICED framework. Furthermore, the minimum CE-MSE solution $p$ ($0\!\leq \!p\! \leq \!1$) can be obtained as the root of a third-order polynomial equation:
\bea
Ap^3+Bp^2+Cp+D=0,
\eea
where $u \!=\!1\!-\!b$, $v \!=\!\frac{\sigma^2}{k}$, and $\rho = \frac{\gamma N_{\text{t}}}{M}$, and
\bea
    A &\!\!\!\!=\!\!\!\!& u^2, \\
    B &\!\!\!\!=\!\!\!\!&-( (\rho\! +\! 2)u^2 \!+\! (1-u)(1+v) ),\\
    C &\!\!\!\!=\!\!\!\!& \rho (u(1\!+\!u)\! +\! (1-u)v) \!-\! (u(1-u) \!+\! (1+u)v \!+\! v^2),  \qquad \\
    D &\!\!\!\!=\!\!\!\!& \rho v ( 2u\!+ \!v ).
\eea
\end{property}

Property~1 provides a theoretical framework to analyse the iterative behaviour of CE and MD on REs with SI-DMRS for general configurations of ($N, M, a, b, N_{\text{t}}$, $\gamma$, $\sigma^2$), enabling the converged CE-MSE and MD-MSE to be determined directly by solving the equilibrium equations. However, to analyse the overall decoding performance across the entire subframe, we must also account for the MD-MSE on REs without DMRS.

\subsection{MD on REs without SI-DMRS}

Once the iterative steps between CE and MD on REs with SI-DMRS have converged, the CE $\tilde{\vec{H}}$ achieves high accuracy and can be utilized for data-detection on the remaining REs\footnote{In AI-based receivers, it is feasible and beneficial to detect data symbols across all REs and leverage all data estimates to further enhance CE by fully exploiting the correlations between the data and channel. This methodology is adopted in our AI receiver design, which is elaborated later.}. On REs without SI-DMRS, the received signal model for detecting $\vec{x}$ reads
\bea \label{data-model}
\vec{y}_{\text{3}} \!&\!\!\!\!=\!\!\!\!&\!\sqrt{\frac{a}{N_{\text{t}}}}(\tilde{\vec{H}}+\vec{\Delta \!H})\vec{x}\!+\!\vec{w} \notag \\
&\!\!\!\!=\!\!\!\!&\sqrt{\frac{a}{N_{\text{t}}}}\tilde{\vec{H}}\vec{x}\!+\!\sqrt{\frac{a}{N_{\text{t}}}}\vec{\Delta\! H}\vec{x}\!+\!\vec{w}. \qquad
\eea

In this case,  the MD-MSE as approximated as $t\vec{I}$ (see Appendix-D), where
\bea
t=  \frac{ap+\sigma^2}{a(1-b)+\sigma^2}.
\eea

\section{Optimal Power Allocations for SI-DMRS Transmissions}

With the CE-MSE and MD-MSE for the ICED process derived above, we now seek to address the following fundamental question: {\textbf{\textit{How can we optimize the power allocation factors $(a, b)$ to maximize the final performance?}}

\subsection{MIESM}
 Before addressing this question, we first introduce the mutual information effective SNR mapping (MIESM) to quantify the combined MD-MSE. The basic principles are as follows:
\begin{itemize}

\item Convert the MD-MSE of the two subsets into effective SNR~\cite{PK08} values:
\bea
\theta_1 = \frac{1}{q}-1, 
\quad
\theta_2= \frac{1}{t}-1.
\eea

\item Map SNR to MI using a constellation-constrained MI function:
\bea \label{mi}
I_i = f_{\mathrm{MI}}\!\left(\frac{\theta_i}{\beta}\right),
\eea
where $f_{\mathrm{MI}}(\cdot)$ is the MI mapping for the chosen modulation and the parameter $\beta$ is a calibration factor. Typical functions of $f_{\mathrm{MI}}(\cdot)$ are shown in Appendix-E.

\item Compute the weighted average MI:
\bea
I_{\mathrm{avg}}=
\frac{M I_1 + N I_2}{M+ N}.
\eea

\item Map the average MI back to an equivalent SNR:
\bea
\theta_{\mathrm{eff}}
=
\beta \, f_{\mathrm{MI}}^{-1}\!\left(I_{\mathrm{avg}}\right),
\eea
\end{itemize}

At last, the combined MD-MSE is calculated as $1/(1\!+\theta_{\mathrm{eff}})$. 

\subsection{Optimal Power Allocations}

\begin{figure}[t]
     \hspace{4mm}
    \centering
    \includegraphics[width=0.55\textwidth]{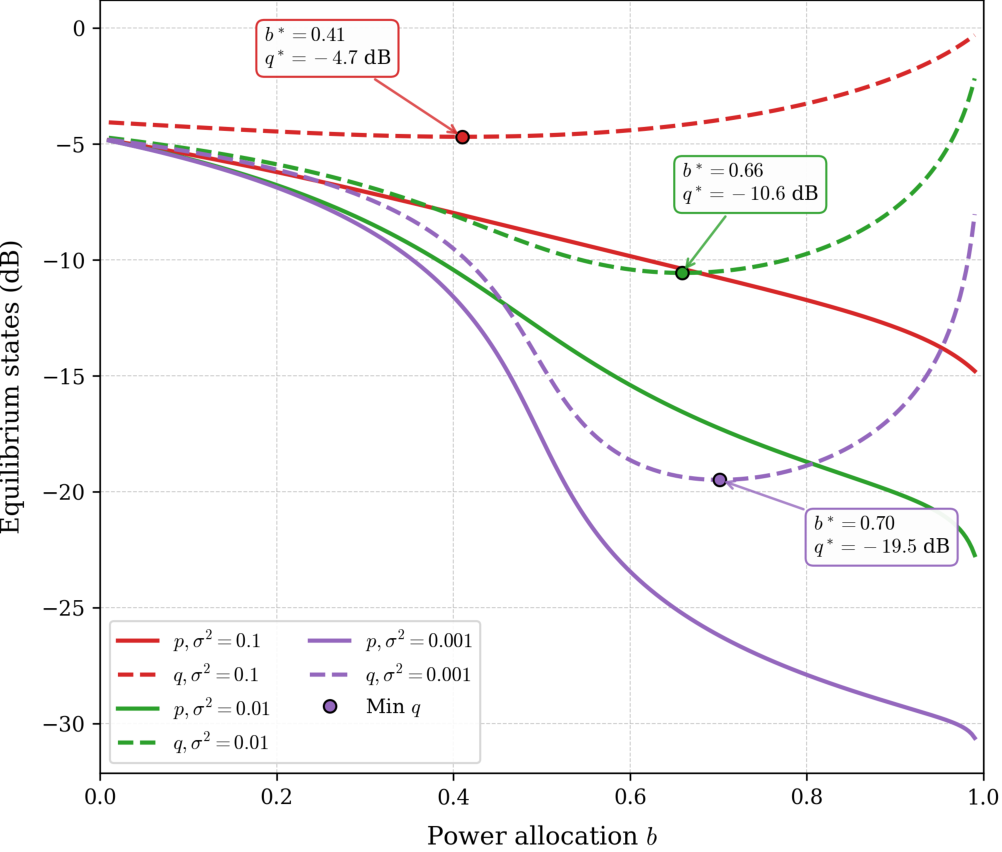}
     \vspace{-4mm}
    \caption{The equilibrium state of CE-MSE and MD-MSE based on Property~1 with ($a\!=\!1$, $\gamma\!=\!2$, $N_{\text{t}}\!=\!4$, $N\!=\!264$, $M\!=\!24$). }
    \label{fig:peqe}
    \vspace{-2mm}
\end{figure}

\begin{figure}
    \hspace{4mm}
    \centering
    \includegraphics[width=0.55\textwidth]{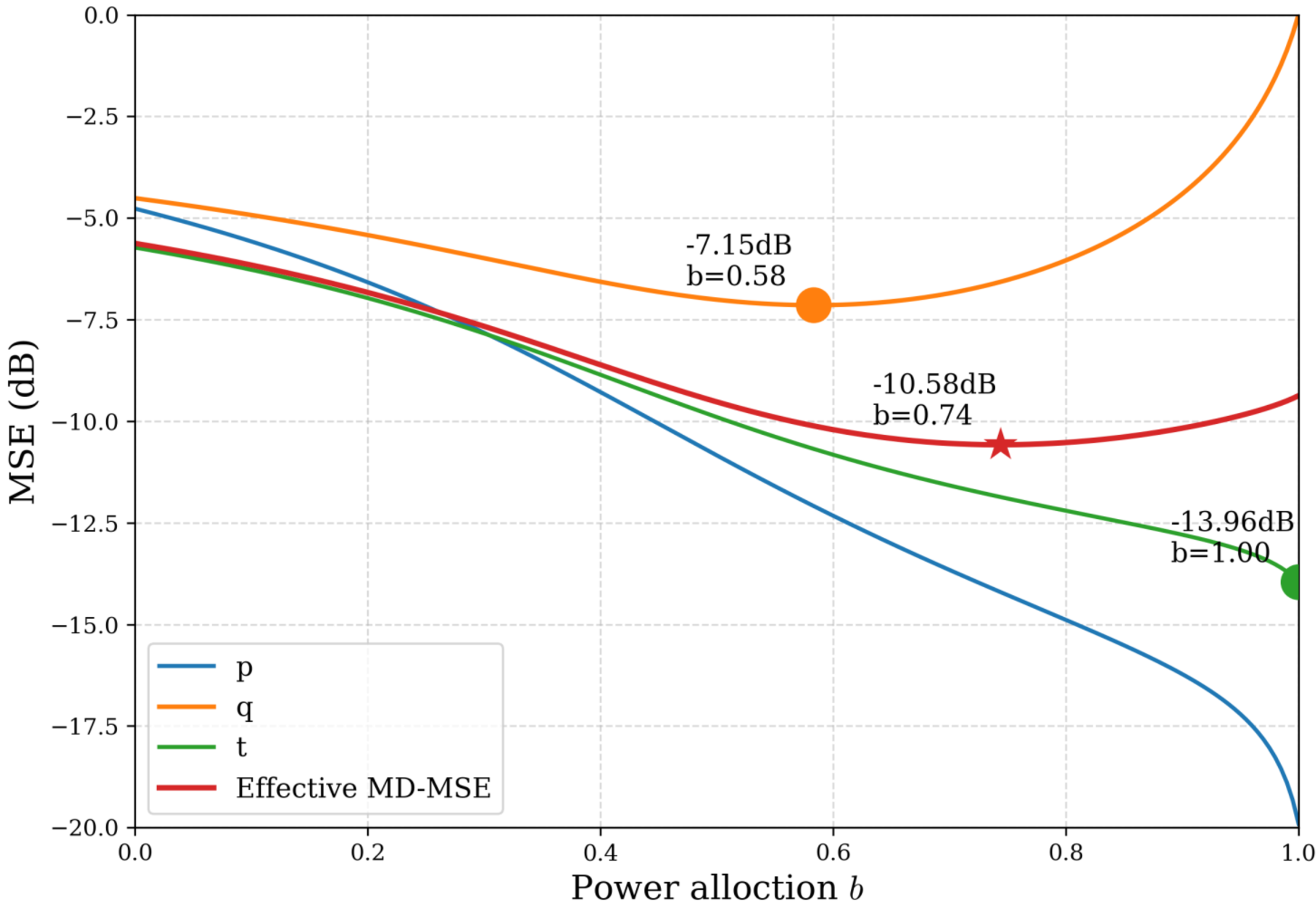}
       \vspace{-4mm}
    \caption{The effective MD-MSE based on MIESM mapping of $p$ and $t$ with ($a\!=\!1$, $\gamma\!=\!2$, $N_{\text{t}}\!=\!4$, $N\!=\!264$, $M\!=\!24$) and $\sigma^2\!=\!-15$dB. }
    \label{fig:MIESM}
     \vspace{-6mm}
\end{figure}

From Property~1, the equilibrium state of $(p, q)$ and $t$ can be directly computed for different power allocation strategies. Subsequently, the effective SNR $\theta_{\mathrm{eff}}$ is obtained from $(q, t)$. Thus, the goal of optimal power allocation is to maximize $\theta_{\mathrm{eff}}$. This provides a clean analytical approach to assess the performance of SI-DMRS, with several illustrative examples shown below.

In Fig.~\ref{fig:peqe}, the values of $p$ and $q$ are illustrated for different noise powers ($\sigma^2 \!=\! -10$dB, $-20$dB, and $-30$dB) with $a \!= \!1$ (i.e., without power-boosting from REs carrying exclusively data symbols). As observed, there exists an optimal value of $b$-the power allocation factor for DMRS-that minimizes $q$ (the MD-MSE), whereas the CE-MSE continues to decrease as $b$ increases. Furthermore, as the SNR increases, the optimal $b$ also increases. This behaviour occurs because, at higher SNRs, the CE-MSE becomes the bottleneck affecting the MD-MSE.

In Fig.~\ref{fig:MIESM}, the effective MD-MSE is presented using $\beta \!=\! 1$ in (\ref{mi}) and $\sigma^2 \!=\! -15$dB, which aligns more closely with overall decoding performance. As seen, the effective MD-MSE-derived via MIESM from ($q$, $t$)-is minimized at a higher value of $b$ compared to the case that considers $q$ in isolation. Furthermore, examining the curve for $t$ indicates that the MD-MSE of data symbols on REs without SI-DMRS requires higher CE accuracy to achieve improved error performance. Note that in both Fig.~\ref{fig:peqe} and Fig.~\ref{fig:MIESM}, the data power allocation factor is fixed to $a \!=\! 1$.

Fig.~\ref{fig:MIESM2D} illustrates the joint optimization over both $a$ and $b$. As demonstrated, the effective MD-MSE is reduced from $-10.42$dB to $-12.2$dB when $a$ is decreased from $1$ to $0.67$. This highlights the significant performance benefits of jointly optimizing power allocation across REs\footnote{A potential drawback of this approach, however, is an increase in the peak-to-average-power ratio (PAPR) for the MIMO-OFDM system.}.

\subsection{Is Superimposing More DMRS with Data Beneficial?}

\begin{figure}[t]
    \hspace{-2mm}
    \centering
    \includegraphics[width=0.6\textwidth]{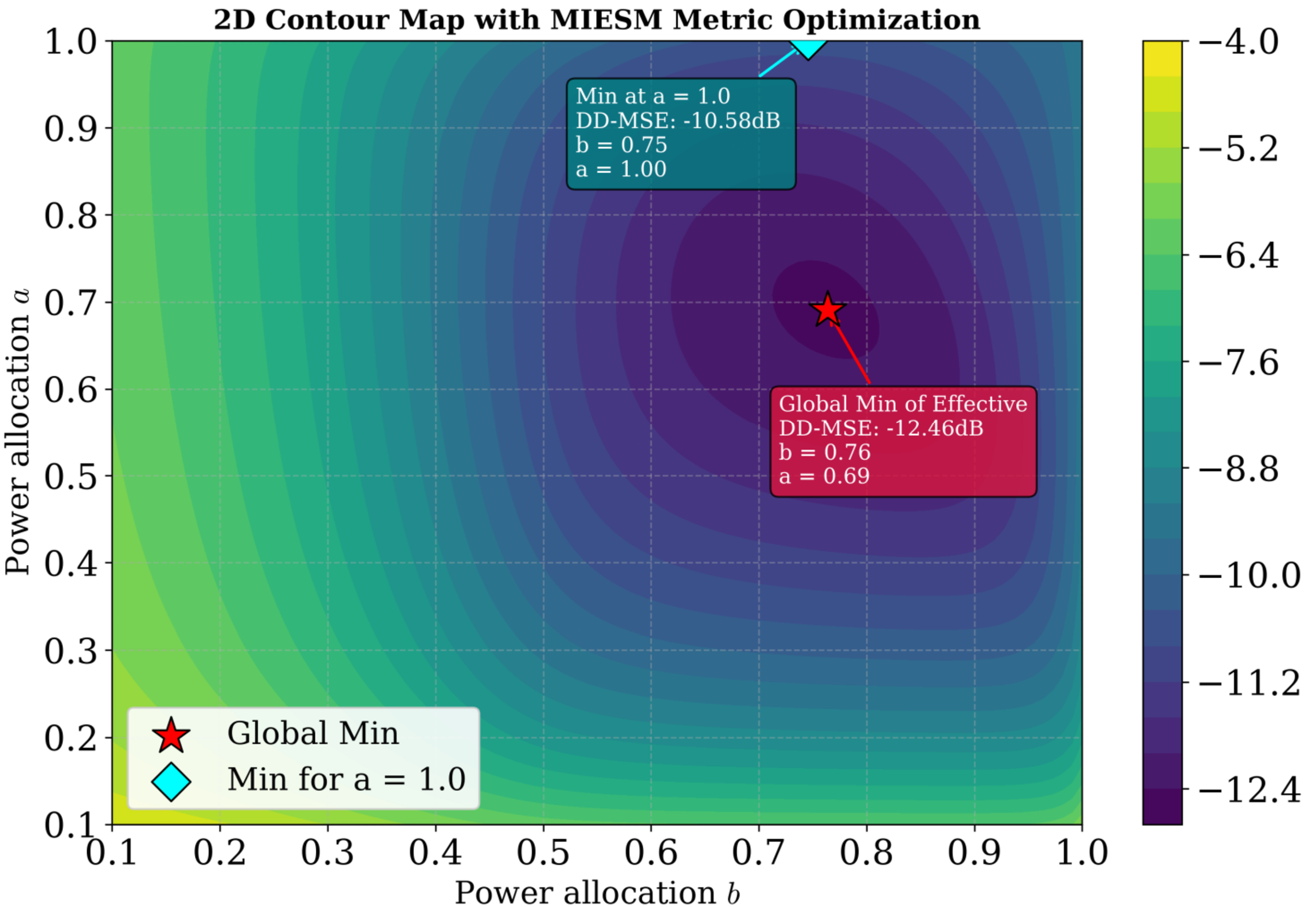}
        \vspace{-4mm}
    \caption{2D contour of effective MD-MSE for joint power allocations over $(a, b)$ with ($\gamma\!=\!2$, $N_{\text{t}}\!=\!4$, $N\!=\!264$, $M\!=\!24$) and $\sigma^2\!=\!-15$dB. }
     \vspace{-6mm}
    \label{fig:MIESM2D}
\end{figure}

\begin{figure}
    \hspace{-4mm}
    \centering
    \includegraphics[width=0.7\textwidth]{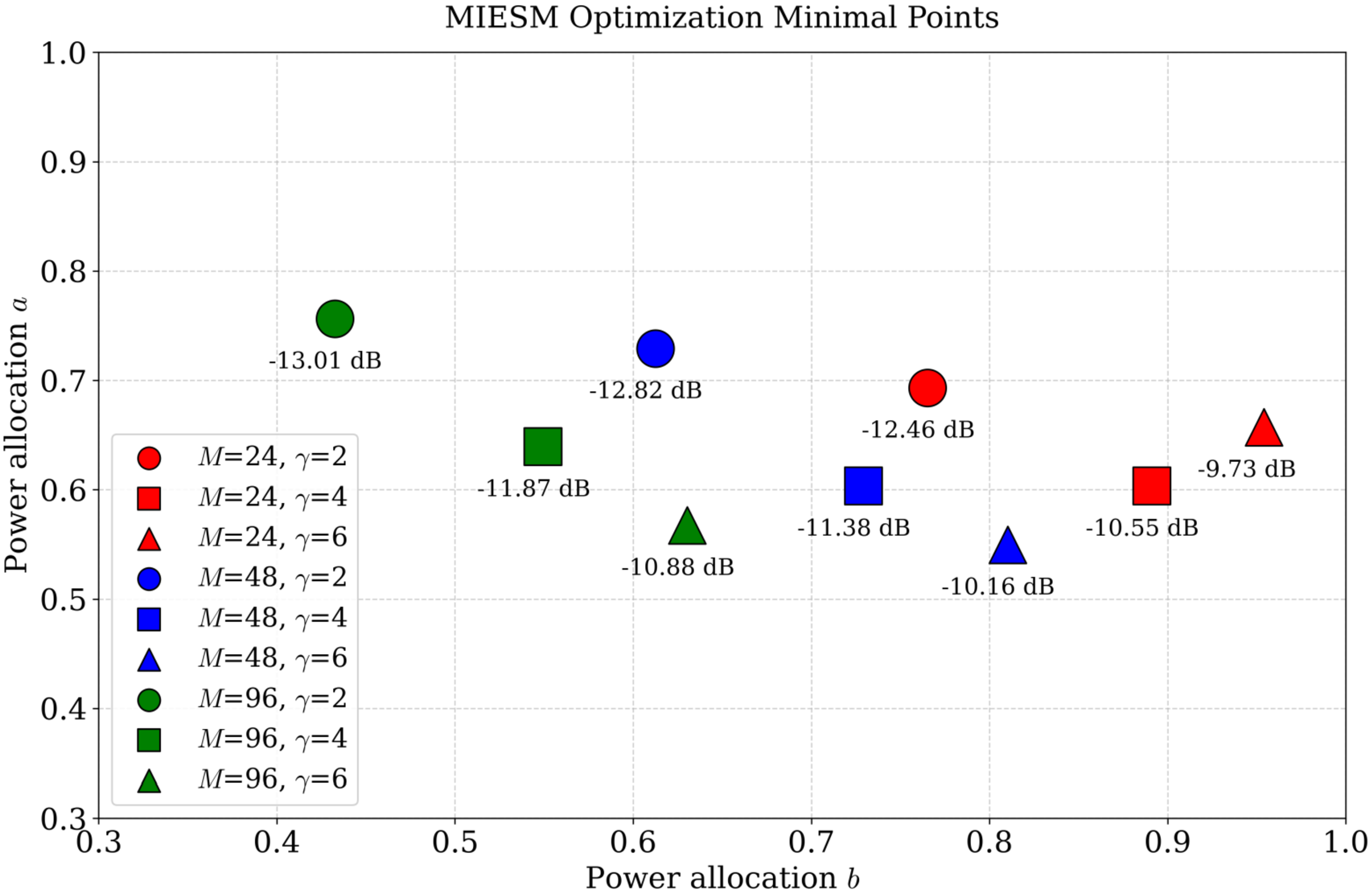}
      \vspace{-4mm}
    \caption{The minimal effective MIESM with different configurations of $M$, i.e., for superimposing more DMRS with data. The gains are limited even under the idea assumption that the de-nosing gain of CE linearly increases in the number of DMRS.}
    \vspace{-6mm}
    \label{fig:pilot_sweep}
\end{figure}

Another interesting question is \textit{\textbf{whether we should superimpose more DMRS with data}}. In the preceding figures, the number of DMRS is set to $M \!= \!24$, with each Tx occupying 6 REs out of a total grid of 240 REs.In Fig.~\ref{fig:pilot_sweep}, we evaluated different configurations of $M$ ($24$, $48$, and $96$) and plotted the minimum effective MD-MSE. For each configuration, we also tested different factors ($\gamma \!= \!2$, $4$, and $6$) representing varying de-noising gains. As shown, for a constant $\gamma \!=\! 2$, increasing $M$ from $24$ to $96$ yields gains. However, keeping $\gamma$ constant implies that the de-noising gain increases linearly with $M$. When compared to the extreme case where the de-noising gain remains unchanged (i.e., when $\gamma/M$ is constant), the configuration with $M\! =\! 24$ and $\gamma \!=\! 2$ outperforms the $M\!= \!48$ and $\gamma \!= \!4$ configuration by $1.55$dB. Therefore, it is not always beneficial to superimpose more DMRS with data, and the optimal choice depends heavily on the de-noising gains and operational system parameters. Nevertheless, this analytical framework provides a valuable perspective for analysing general cases to optimize SI-DMRS.

\section{The Proposed Transformer based AI-ICED Receiver Design}

With the theoretical framework established, we next present the AI-ICED receiver design, which harnesses the spectral efficiency (SE) gains enabled by SI-DMRS transmission. The overall architecture is illustrated in Fig.~\ref{fig:system_model}. The AI receiver processes three tensor inputs: the received signal tensor $\mathbf{Y}$, the DMRS symbols $\mathbf{s}_{\mathrm{p}}$, and the DMRS coordinate mask $\Omega_{\mathrm{p}}$. By treating the DMRS information as contextual prompts, the AI architecture establishes a direct mapping from the observation space to the symbol space. Meanwhile, the AI receiver applies an ICED design, and at iteration stage $i$, it updates the CE and the a posteriori probabilities of symbol vectors, $p^{i}(\vec{x} \!\mid\! \vec{Y}, \tilde{\vec{H}}^{(i-1)}, \tilde{\vec{x}}^{(i-1)})$, by capturing the dependencies between the intermediate CE and soft symbol estimates from the previous iteration. Ultimately, the logits generated by the AI-ICED receiver are used to calculate bit log-likelihood ratios (LLRs), which are then fed into the channel decoder. As depicted in Fig.~\ref{fig:system_model}, the overall design is structured as a cascaded loop where the inputs at each stage are processed through three primary modules: the tokenizer, the ICED module, and the physics-guided feature construction (PGFC) module between iteration stages.

\begin{figure*}[t]
    \centering
    \includegraphics[width=1\textwidth]{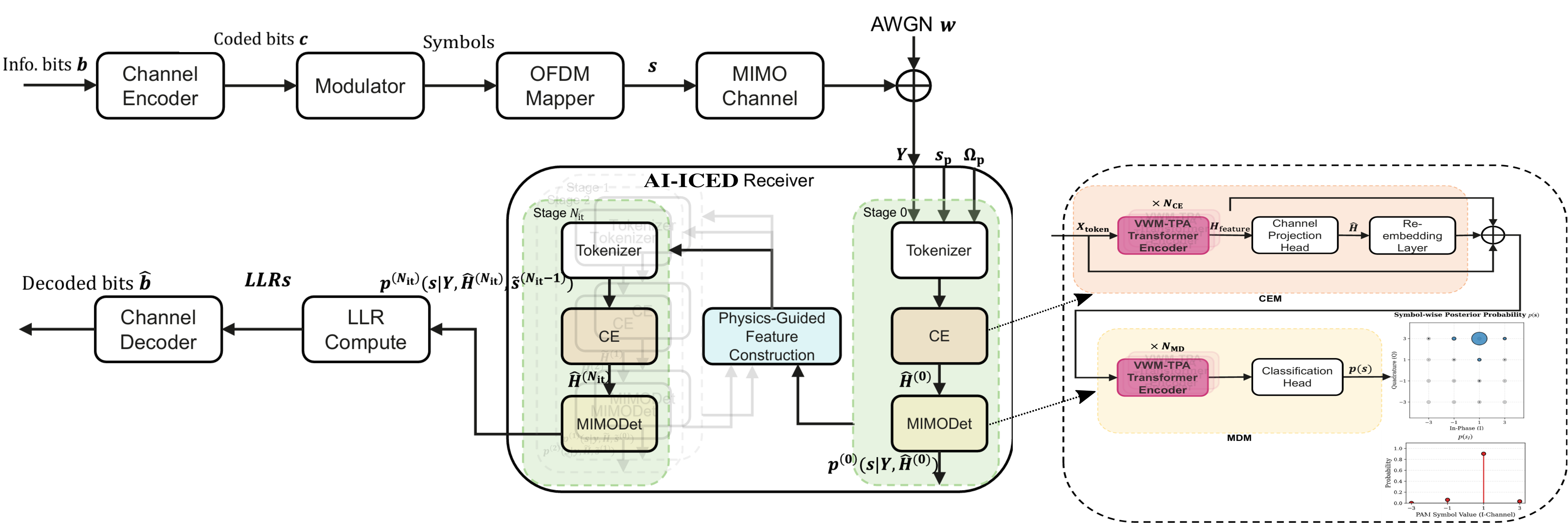}
        \vspace{-12mm}
    \caption{The structure of the MIMO-OFDM system and the proposed AI-ICED receiver, along with details regarding the NN construction and configurations provided in Appendix F through I and Table III.}
    \label{fig:system_model}
        \vspace{-6mm}
\end{figure*}

\subsection{Tokenization}

This module serves as a tensor pre-processor that performs a spatial flattening operation, concatenates the tensors along their feature axes, and applies a linear projection. This step maps the physical tensors into a unified $d_{\mathrm{virtual}}$-dimensional token sequence. Note that all defined complex-valued tensors include a batch dimension $B$, with the real and imaginary components decoupled and stacked along a dedicated feature axis of size two. The module receives the real-valued observation tensor $\vec{Y}_{\mathbb{R}} \!\in\! \mathbb{R}^{B \times N_{\mathrm{r}} \times N_{\mathrm{sym}} \times N_{\mathrm{sc}} \times 2}$, the symbol tensor $\vec{s}_{\mathbb{R}}\! \in\! \mathbb{R}^{B \times N_{\mathrm{t}} \times N_{\mathrm{sym}} \times N_{\mathrm{sc}} \times 2}$, and the binary mask $\vec{\Omega}_{\mathrm{p}}\! \in\! \{0, 1\}^{B \times N_{\mathrm{t}} \times N_{\mathrm{sym}} \times N_{\mathrm{sc}}}$ indicating where an SI-DMRS is transmitted.

The raw feature vector at a specific RE coordinate $(n, k)$ is constructed by concatenating the flattened antennas and the real-imaginary components, and the mapping is formulated as
\be \label{xraw}\vec{x}_{\mathrm{raw}}^{(i)}[n, k] = \operatorname{Concat} \big(\vec{\Phi}^{(i-1)}[n, k], \tilde{\vec{s}}^{(i-1)}[n, k], \vec{\Omega}_{\mathrm{p}}[n, k] \big),\ee
where the symbol vector is reconstructed according to the SI-DMRS scheme and power allocations as
\bea
    \tilde{\vec{s}}^{(i)}_{n,k} = 
    \begin{cases} 
    \sqrt{\rho} \, s_{\text{DMRS}} \!+\! \sqrt{1 - b} \, \tilde{\vec{x}}^{(i-1)}_{n,k}, & \text{if }  \vec{\Omega}_{\mathrm{p}}(n,k)=1 \\ 
    a\tilde{\vec{x}}^{(i-1)}_{n,k}, & \text{Otherwise}
    \end{cases}\!. 
\eea

At initial stage ($i\!=\!0$), the tensor $\vec{\Phi}^{(-1)}[n, k]$ is initialized by $\vec{Y}_{\mathbb{R}}[n, k]$, and $\tilde{\vec{x}}^{(-1)}[n, k]\!=\!\vec{0}$ such that $\tilde{\vec{s}}^{(0)}_{n,k}$ only contains known DMRS. The dimension of $\vec{x}_{\mathrm{raw}}^{(0)}[n, k]$ is $D_{\mathrm{in}}^{(0)} \!= \!2N_{\mathrm{r}} \!+\! 3N_{\mathrm{t}}$. In subsequent stages, the dimension $D_{\mathrm{in}}^{(i)}$ expands to accommodate more features generated by the PGFC module. 

A projection matrix $\mathbf{W}^{(i)} \!\in\! \mathbb{R}^{D_{\mathrm{in}}^{(i)} \times d_{\mathrm{virtual}}}$ is applied to map the raw feature vector into a $d_{\mathrm{virtual}}$-dimensional latent space as $\vec{x}_{\mathrm{raw}}^{(i)}[n, k]\mathbf{W}^{(i)}$. By aggregating all projected vectors across $L$ REs, the tokenizer outputs a sequence $\vec{X}_{\mathrm{token}}^{(i)}\! \in\! \mathbb{R}^{B \times L \times d_{\mathrm{virtual}}}$ as input to the next ICED module. This tokenization strategy closely aligns with the physical characteristics of MIMO-OFDM systems. By flattening the 2D time-frequency grids into a sequence of length $L$, the self-attention mechanism is augmented with Rotary Position Embedding (RoPE)~\cite{su24} and effectively captures the local coherence of MIMO channels. Furthermore, by collapsing spatial antennas into a single $d_{\mathrm{virtual}}$-dimensional feature vector per token, the architecture delegates the resolution of spatial correlations to the network. This structural design enables the attention layers to perform MD internally within the high-dimensional latent representation.

\begin{figure}[t]
    \centering
    \includegraphics[width=0.27\textwidth]{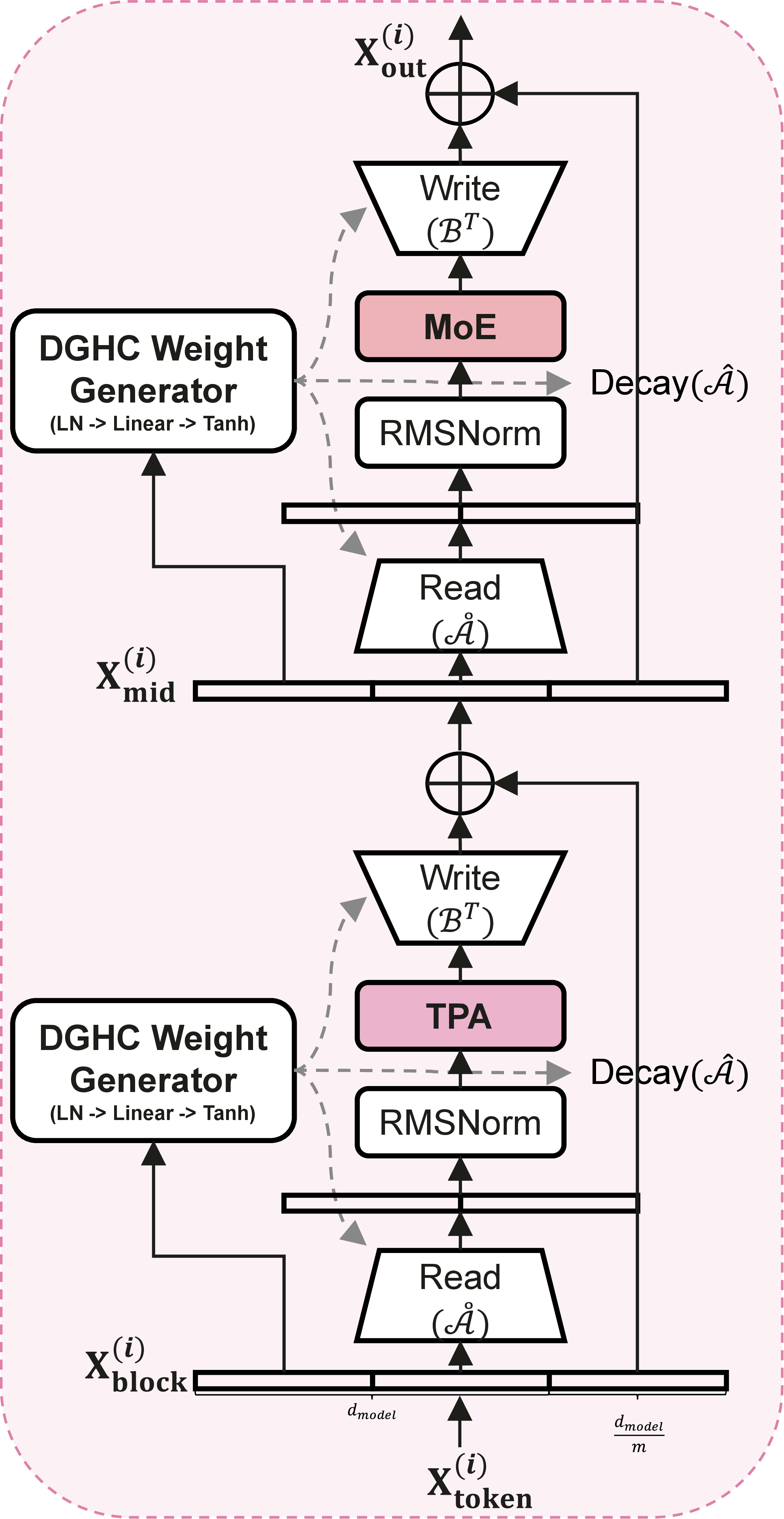}
         \vspace{-4mm}
    \caption{The VWN-TPA-MoE architecture enhanced Transformer encoder design.}
     \vspace{-6mm}
    \label{fig:vwn}
\end{figure}

\subsection{ICED Backbone}

The ICED operating on the tokenized sequence is designed as an unfolded cascaded architecture that performs CE and MD. Rather than treating them as disjoint modules, the AI-ICED design unifies them into a fused pipeline, as illustrated in Fig.~\ref{fig:system_model}. The Transformer encoder architecture incorporates with the latest VWN, TPA and MoE techniques to optimize feature representation.

The prompt $\mathbf{X}_{\mathrm{token}}^{(i)}$ received from the tokenizer is firstly processed by $N_{\mathrm{CE}}$ stacked customized Transformer encoder blocks that incorporates VWN, TPA, and MOE. These specialized encoder blocks iteratively resolve the time-frequency correlations and spatial interference, yielding a channel latent representation $\vec{H}_{\mathrm{feature}}^{(i)} \!\in\! \mathbb{R}^{B \times L \times d_{\mathrm{virtual}}}$. To reconstruct the MIMO channel, a projection head (a linear fully-connected layer) $\mathbf{W}_{\mathrm{CE}}^{(i)}$ maps $\mathbf{H}_{\mathrm{feature}}^{(i)}$ back to physical antenna dimensions
\bea
    \hat{\vec{H}}_{\mathrm{flat}}^{(i)} = \vec{H}_{\mathrm{feature}}^{(i)} \vec{W}_{\mathrm{CE}}^{(i)} \in \mathbb{R}^{B \times L \times (2 N_{\mathrm{r}} N_{\mathrm{t}})}.
\eea
It is subsequently reshaped to the real-valued CE tensor $\hat{\vec{H}}^{(i)} \! \in\! \mathbb{R}^{B \times N_{\mathrm{r}} \times N_{\mathrm{t}} \times N_{\mathrm{sym}} \times N_{\mathrm{sc}} \times 2}$.

Instead of solely using $\hat{\mathbf{H}}_{\mathrm{flat}}$, the input to MD module is formulated via an additive residual bridge
\bea \label{eq:fusion_bridge}
    \vec{Z}_{\mathrm{dd}}^{(i)} = \vec{X}_{\mathrm{token}}^{(i)} + \vec{H}_{\mathrm{feature}}^{(i)} + \hat{\vec{H}}_{\mathrm{flat}}^{(i)} \vec{W}_{\mathrm{re-embed}}^{(i)},
\eea
where $\vec{W}_{\mathrm{re-embed}}^{(i)}$ acts as a re-embedding layer that projects the CE back into the high-dimensional latent space. This design linearly aggregates the original input prompt $\mathbf{X}_{\mathrm{token}}^{(i)}$, the uncompressed high-dimensional channel memory $\mathbf{H}_{\mathrm{feature}}^{(i)}$, ensuring maximum information retention\footnote{However, the purpose of channel projection head and re-embedding is just to obtain the estimate of $\vec{H}$ and measure the CE-MSE for training loss optimization during training, once the NN is trained, these two modules can be combined together as one in inference stage.}.

The fused sequence $\mathbf{Z}_{\mathrm{dd}}^{(i)}$ is subsequently processed by $N_{\mathrm{MD}}$ transformer encoder blocks, which apply the same VWN-TPA-MOE design to detect the symbols. Finally, a classification head maps the output features into logits for the real and imaginary pulse amplitude modulation (PAM) symbols:
\begin{equation}
    \vec{Z}_{\mathrm{PAM}}^{(i)} = \mathrm{Encoder}_{\mathrm{MD}}\big(\vec{Z}_{\mathrm{dd}}^{(i)}\big) \vec{W}_{\mathrm{MD}}^{(i)} ,
\end{equation}
where $\vec{W}_{\mathrm{MD}}^{(i)}$ is the projection weight, and $\sqrt{M}$ represents the PAM alphabet size converted from the $M$-QAM modulation. The output tensor $\vec{Z}_{\mathrm{PAM}}^{(i)}\! \in\! \mathbb{R}^{B \times L \times (2 N_{\mathrm{t}} \sqrt{M})}$ encompasses the real and imaginary logits, $\vec{Z}_{\Re}^{(i)}$ and $\mathbf{Z}_{\Im}^{(i)}$, which are utilized for loss computation. By applying standard softmax normalization, these logits are mapped to the probability tensor $\hat{\vec{P}}^{(i)}$.

\subsection{Physics Guided Feature Construction (PGFC)}

While the VWN-TPA-MoE encoders provide a powerful computational engine, a purely data-driven approach faces limitations in modeling the high-dimensional state space of MIMO-OFDM detection. Rather than propagating raw outputs across refinement stages, PGFC functions as a differentiable bridge. It constructs communication-theoretic tensors derived from current channel and symbol estimates and supplies the subsequent VWN-TPA encoder with structured features. This relieves the network from learning fundamental spatial projections strictly from data, thereby accelerating convergence and improving generalization across varying channel conditions.

Using the probability tensor $\hat{\vec{P}}^{(i)}$ output from the ICED module, the soft estimates $\tilde{\vec{x}}^{(i)}$ of the transmitted data can be computed. Based on these estimates, several features are assembled for the next stage: an interference cancellation (IC) result, $\vec{R}^{(i)} = \vec{Y} - \hat{\vec{H}}^{(i)}\tilde{\vec{S}}^{(i)}$; matched-filter (MF) outputs, $\vec{Z}_{\mathrm{MF}}^{(i)} = (\hat{\vec{H}}^{(i)})^{\mathsf{H}} \vec{Y}$ and $\vec{R}_{\mathrm{MF}}^{(i)} = (\hat{\vec{H}}^{(i)})^{\mathsf{H}} \vec{R}^{(i)}$; and the Gram matrix $\vec{G}^{(i)} = (\hat{\vec{H}}^{(i)})^{\mathsf{H}} \hat{\vec{H}}^{(i)}$.
After converting them into their real-valued versions, the tensor $\vec{\Phi}^{(i)}$ in (\ref{xraw}) sent to the tokenizer is augmented as follows:
\bea 
    \vec{\Phi}^{(i)} = \operatorname{Concat}\bigl( \vec{Y}_{\mathbb{R}},\; \vec{R}_{\mathbb{R}}^{(i)},\; \vec{Z}_{\mathrm{MF},\mathbb{R}}^{(i)},\; \vec{G}_{\mathbb{R}}^{(i)},\; \vec{R}_{\mathrm{MF},\mathbb{R}}^{(i)} \bigr).
\eea

Combined with the updated soft state $\tilde{\mathbf{s}}^{(i)}$ and the pilot mask $\vec{\Omega}_{\mathrm{p}}$ are forwarded to the tokenizer for the next ICED stage\footnote{A stop-gradient operation is applied to $\tilde{\vec{s}}^{(i)}$ to isolate the backward pass within each iterative stage, stabilizing the deep supervision dynamics across the unfolded cascade}.

\subsection{Training Loss Design}

To simultaneously optimize CE and MD, the training loss at iteration stage $i$ balances these objectives using a hyper-parameter $\beta_1 \in [0, 1]$:
\bea 
\mathcal{L}_{\text{stage}}^{(i)} = \beta_1 \mathcal{L}_{\vec{H}}^{(i)} + (1 - \beta_1) \mathcal{L}_{\vec{S}}^{(i)},
\eea
where the detection loss is measured via cross-entropy as
\bea
\mathcal{L}_{\vec{S}}^{(i)} = \mathrm{CrossEntropy}\big(\vec{Z}_{\text{PAM}}^{(i)}, \vec{Z}_{\text{PAM},\text{Label}}\big),
\eea
and the CE loss is computed using the MSE:
\bea
\mathcal{L}_{\vec{H}}^{(i)} = \mathrm{MSE}\big(\hat{\vec{H}}^{(i)}, \mathbf{H}_{\text{Label}}\big).
\eea 

To further mitigate vanishing gradients across the unfolded cascade, deep supervision injects gradients into all intermediate stages. Controlled by a weighting factor $\beta_2 \in [0, 1]$, the total loss over $N_{\mathrm{it}}\!+\! 1$ stages is defined as
\bea
    \mathcal{L}_{\mathrm{total}} = (1 - \beta_2) \mathcal{L}_{\mathrm{stage}}^{(N_{\mathrm{it}})} + \frac{\beta_2}{N_{\mathrm{it}}} \sum_{i=0}^{N_{\mathrm{it}} - 1} \mathcal{L}_{\mathrm{stage}}^{(i)}.
\eea

\subsection{Inference Complexity}

A critical challenge for AI-based receivers is inference complexity. The per-stage computational complexity is primarily bounded by attention-score computation, low-rank TPA projections, and MoE operations. As the unfolded receiver iterates over $N_{\mathrm{it}} \!+\! 1$ stages, the total inference cost scales linearly with the number of applied stages .To manage this complexity, the architecture incorporates three key design strategies: VWN expands the latent memory capacity to $d_{\mathrm{virtual}}\! =\! (n/m)d_{\mathrm{model}}$, but restricts dense nonlinear operations to $m$ out of $n$ virtual slots. This effectively decouples representational capacity from active computational width. TPA factorizes the Query, Key, and Value projections into low-rank tensor products. Given a query rank $r_q$ and a shared key/value rank $r$, the dense projection complexity drops from $\mathcal{O}(L d_{\mathrm{model}}^2)$ to $\mathcal{O}(L d_{\mathrm{model}}(r_q \!+\!  2r))$. Meanwhile, the $\mathcal{O}(L^2 d_{\mathrm{model}})$ attention-score term is retained to capture global time-frequency dependencies. The MoE feed-forward block expands parameter capacity via expert specialization. By activating exactly $N_s$ shared experts and $K_r$ routed experts per token, the active computation maintains FLOP equivalence with a standard dense Transformer FFN. A list of the complexity is summarized in Table~I in Appendix~I.

\section{Numerical Results}

Next we analyse the performance of the AI-ICED receiver under SI-DMRS and 5G-NR settings. The configurations and channel models are aligned with 3GPP specifications and are summarized in Table~\ref{tab:sys_params}, and the hyper-parameter configurations for the AI-ICED receiver construction are summarized in Table~3, respectively. The basic configurations are the same as depicted in Fig.~1, where we assume a bandwidth of two PRBs, and two OFDM symbols are used for SI-DMRS transmissions (so $M\!=\!48$), regardless of $N_{\text{t}}$, and $N\!=\!240$. So the maximal SE gain can be obtained is 20\%. Performance is characterized in terms of CE-MSE, BLER, and throughput for different SNRs and code-rates. For all SI-DMRS configurations, the DMRS pattern adopts the layout shown in Fig.~\ref{fig:pilot_strategies}. Based on the theoretical results in Sec.~III, which are also validated by our tests, the power allocation factor is set to $b \!=\! 0.8$, assigning a larger proportion of the transmit power to the pilot symbols. For a fair comparison with the 5G-NR baseline, no power boosting is applied (i.e., $a \!=\! 1$). 

The multipath fading channels are modelled as Extended Typical Urban (ETU70)~\cite{3gpp36101}. For evaluation, the generated datasets comprise 800,000 frames for the $2 \!\times\! 2$ MIMO setup, alongside 200,000  subframes for the $4 \!\times\! 4$ MIMO configuration. All datasets are partitioned into 81\% for training, 9\% for validation, and 10\% for testing. Finally, the block error rate (BLER) and throughput of the proposed architecture are benchmarked against classical baselines including conventional LMMSE and the optimal log-map based maximum likelihood detector (MLD). Furthermore, to decouple the errors induced by CE from detection, a genie-aided variant of the proposed architecture is tested. It bypasses the estimated channel tensor $\hat{\vec{H}}_{\text{flat}}$ and directly injects the ground-truth $\vec{H}$ into the latent residual fusion bridge, which reflects the MIMO detection performance under genie-aided channel state information (CSI).

Note that the same AI-ICED design is also applied to the 5G-NR baseline simulations using separate training sessions, which corresponds to a special case of SI-DMRS where $b\! = \!1$. Consequently, the overarching design principles and AI receiver architecture apply equally to standard 5G-NR configurations.

\subsection{CE-MSE and BLER Convergences of the Proposed AI-ICED Receiver}

   \begin{figure}[t]
        \centering
        \includegraphics[width=0.6\textwidth]{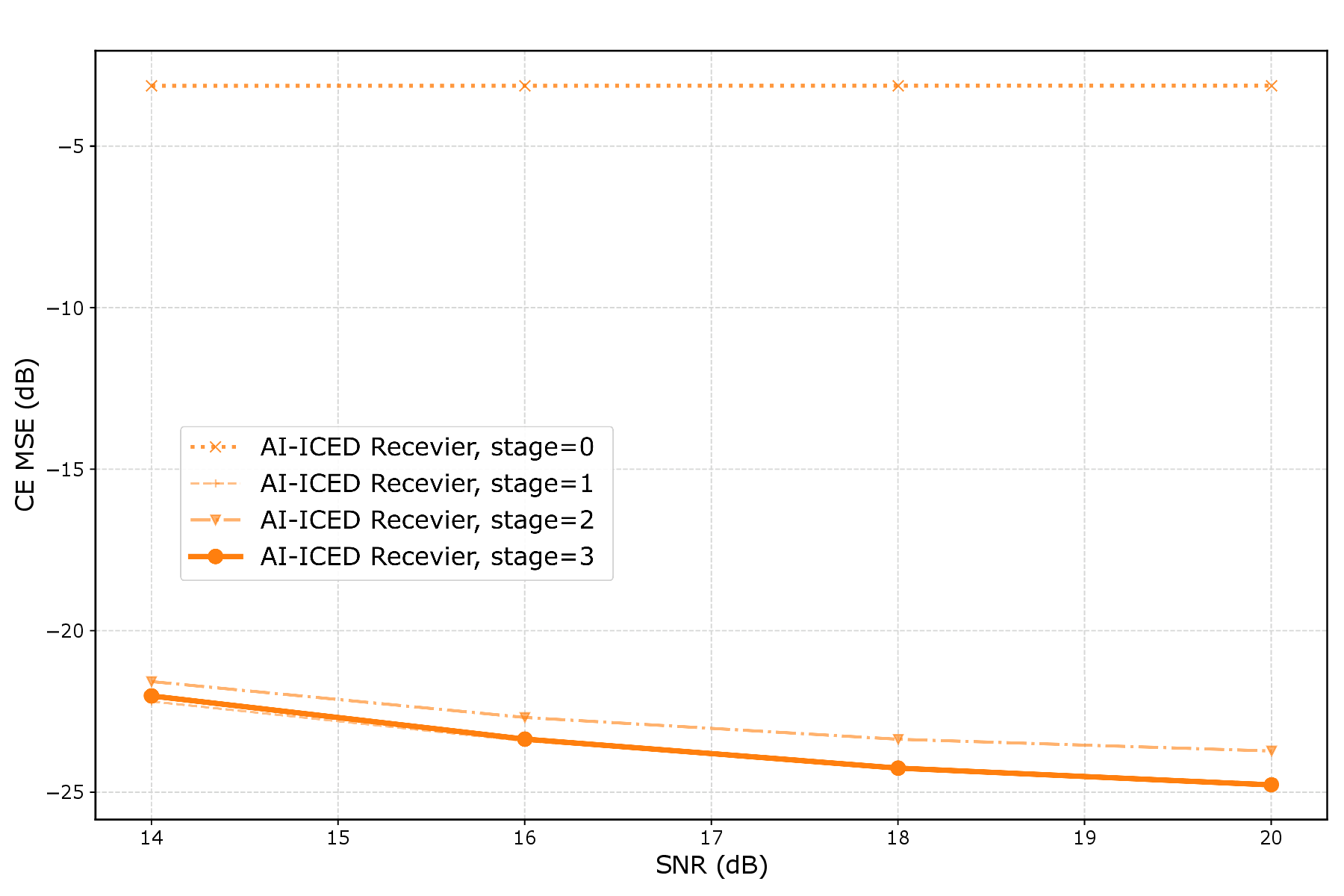}
        \vspace{-6mm}
        \caption{CE-MSE over iteration stages for  $4 \!\times\! 4$ MIMO and 16QAM modulation.}
        \label{fig:it_stage_loss_h_si}
        \vspace{-3mm}
    \end{figure}
    
    \begin{figure}
        \centering
        \includegraphics[width=0.6\textwidth]{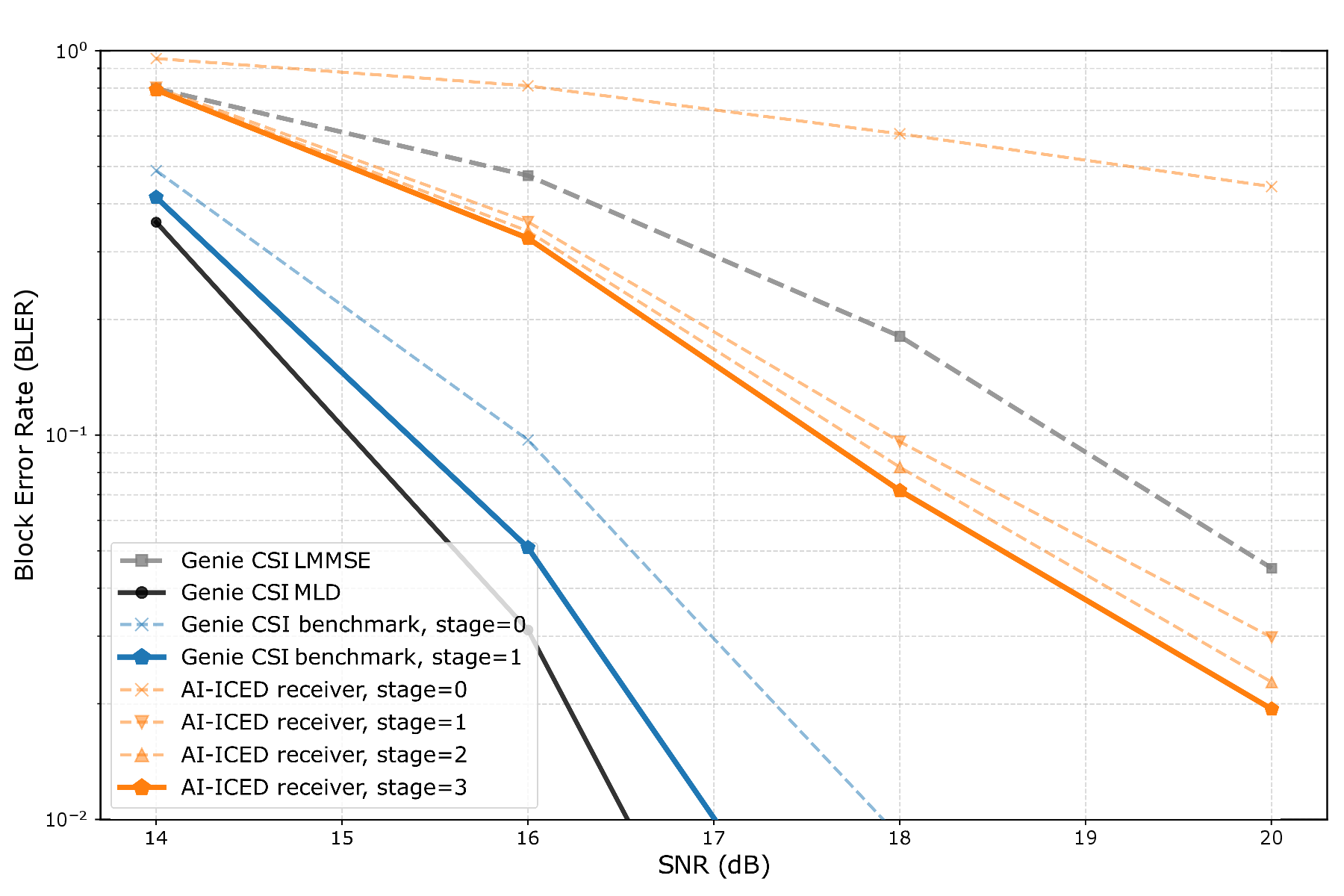}
         \vspace{-6mm}
        \caption{BLER performance with the same configurations in Fig.~8.}
        \label{fig:it_bler_si}
        \vspace{-6mm}
    \end{figure}

\begin{figure}
    \centering
    \includegraphics[width=0.6\textwidth]{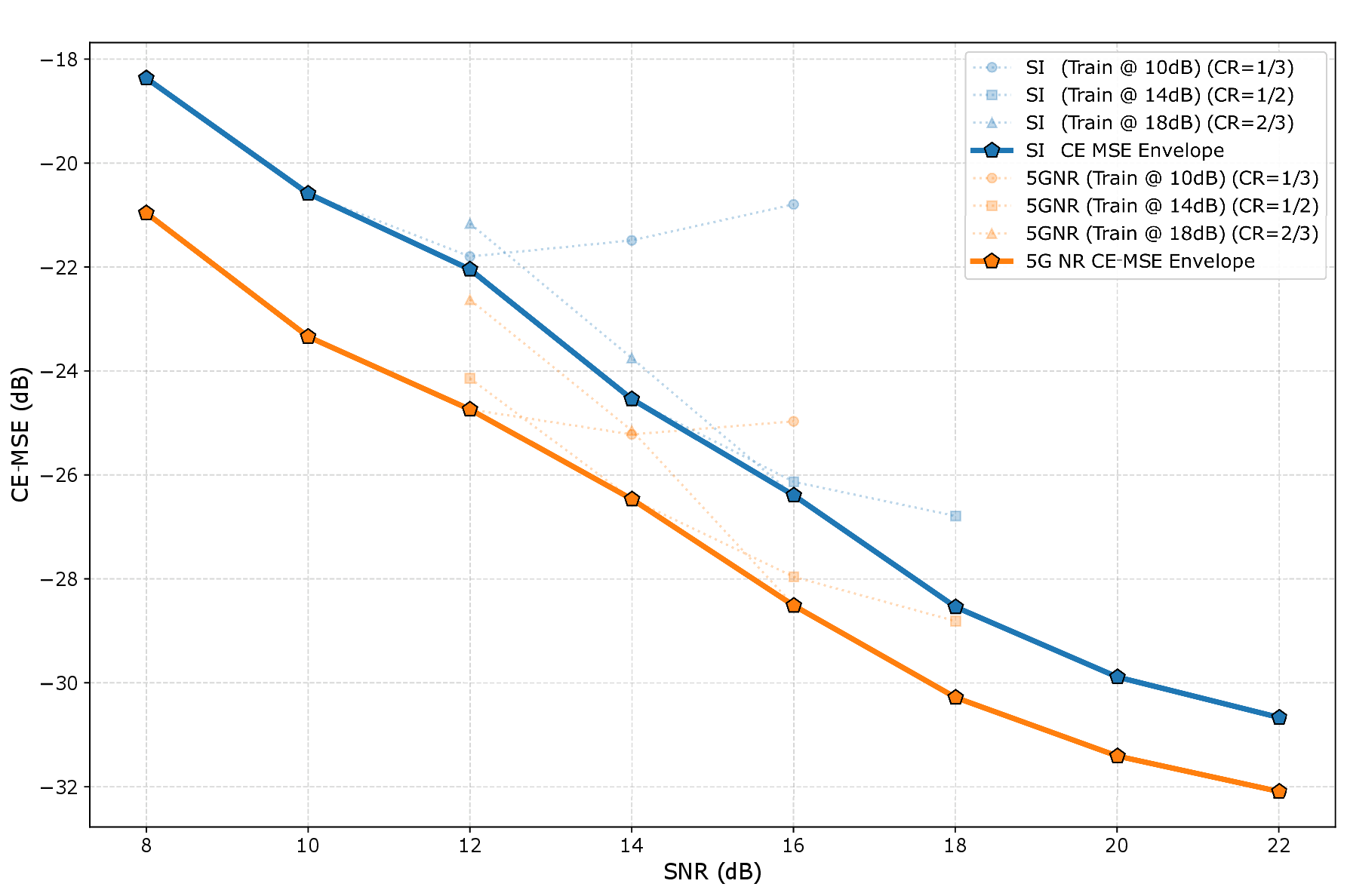}
    \vspace{-6mm}
    \caption{CE-MSE under $2 \!\times \!2$ MIMO and 16-QAM across a wide SNR range.}
    \label{fig:si_ce_mse_22}
    \vspace{-2mm}
\end{figure}

The CE-MSE and BLER performance of the proposed AI-ICED receiver under a $4 \!\times\! 4$ MIMO and 16-QAM modulation (4-PAM) are presented in Fig.~8 and Fig.~9, respectively, using an LDPC code-rate of $2/3$. The AI-ICED receiver is trained at an SNR of $16$dB, but evaluated across a wide SNR range. As shown, both the CE-MSE and BLER converge rapidly within a single iteration. With two additional iterations, the performance gain is approximately $0.5$dB in SNR for both metrics, highlighting the effectiveness of the iterative AI-ICED design. Furthermore, as illustrated in Fig.~9, the AI-ICED receiver approaches the performance of the optimal MLD provided with genie-aided CSI.

In Fig.~\ref{fig:si_ce_mse_22}, the CE-MSE across a wide SNR range is presented, along with the corresponding required operational SNR for different code-rates. Although the AI-ICED receiver trained at a fixed SNR can generalize to adjacent SNRs, train-test SNR mismatch prevents optimal detection over the entire operating range. To handle this mismatch, dedicated models are trained at regular SNR intervals to capture variations and maintain optimality. As shown, due to the interference introduced between DMRS and data symbols under SI-DMRS, there is a CE-MSE degradation of approximately $2\text{--}4\text{ dB}$ compared to the 5G-NR baseline. However, the CE-MSE error remains insignificant compared to the noise power. Furthermore, this performance loss is well justified by the increased SE and higher throughput achieved by the AI-ICED receiver as shown later.

\subsection{Throughput Increments}

\begin{figure}[t]
    \centering
    \includegraphics[width=0.6\textwidth]{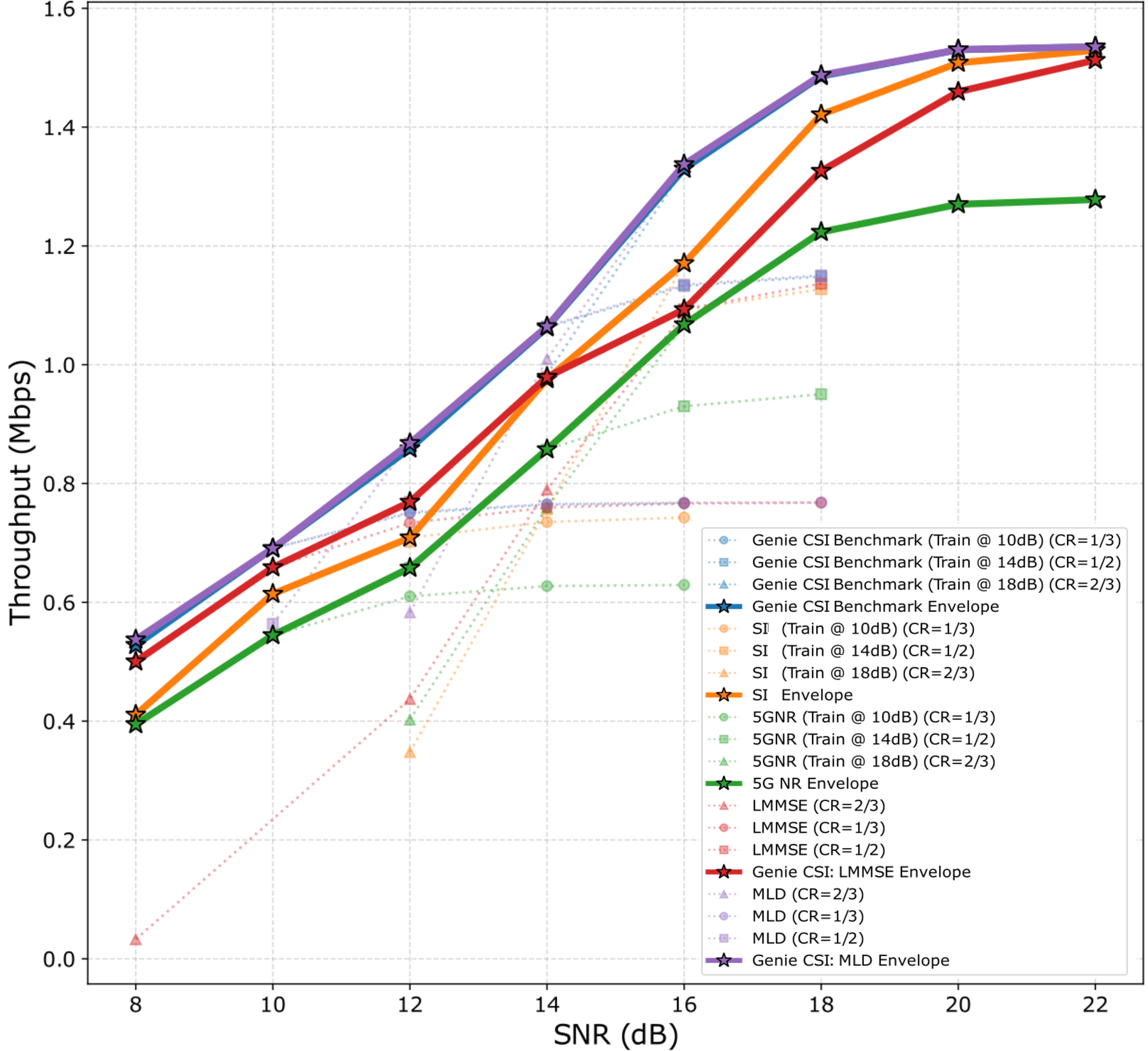}
    \vspace{-6mm}
    \caption{Throughput envelopes under the $2\! \times\! 2$ and 16-QAM modulation.}
    \vspace{-6mm}
    \label{fig:si_throughput_22}
\end{figure}

\begin{figure}
    \centering
    \includegraphics[width=0.6\textwidth]{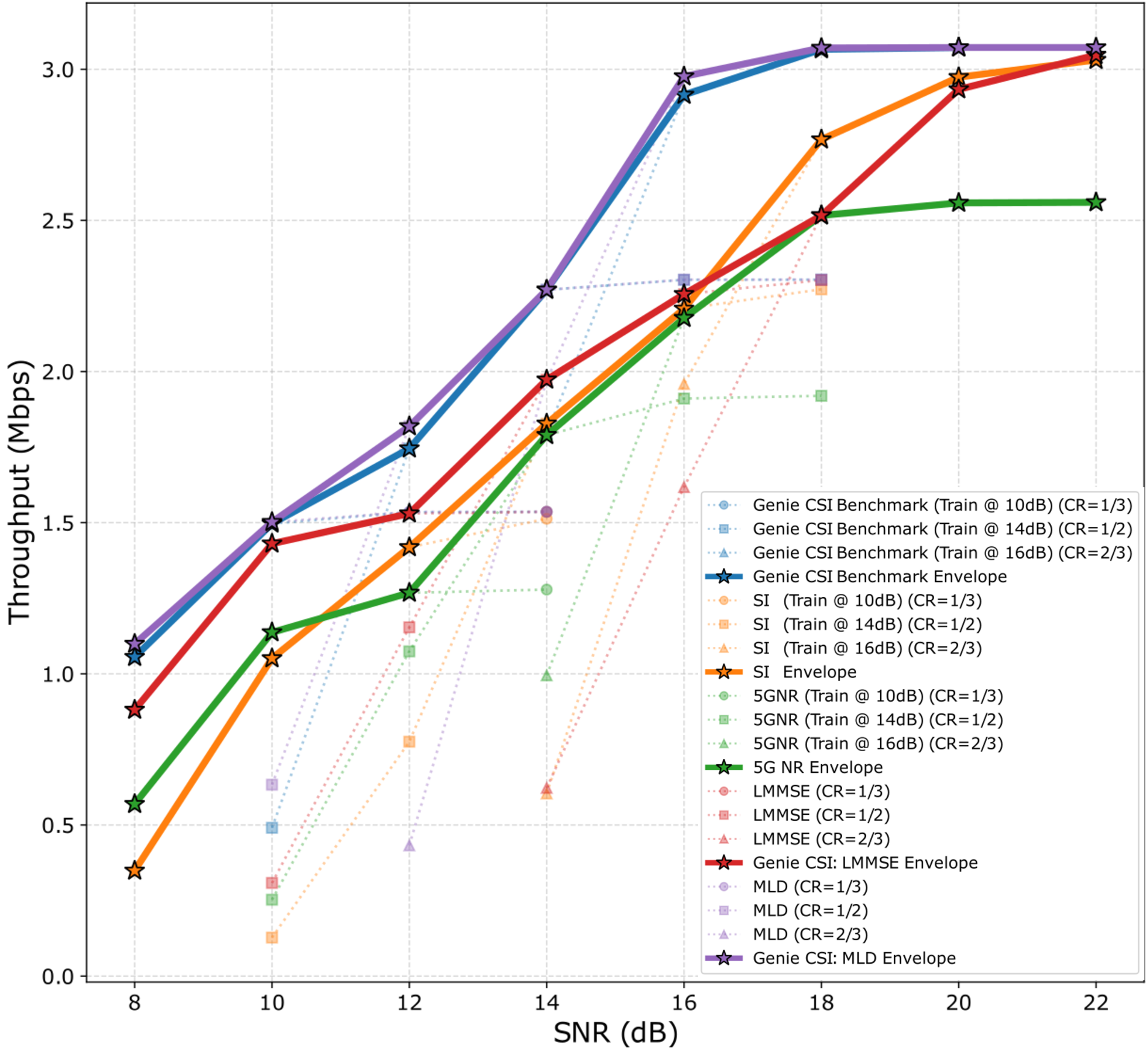}
    \vspace{-6mm}
    \caption{Throughput envelopes under the $4\! \times\! 4$ and 16-QAM modulation.}
    \label{fig:si_throughput_44}
    \vspace{-6mm}
\end{figure}

Although both the CE-MSE and BLER can be worse than 5G-NR baselines without superimposed DMRS, the primary advantage of SI-DMRS is the increased SE resulting from its zero DMRS overhead. The throughput envelopes are illustrated in Fig.~\ref{fig:si_throughput_22} and Fig.~\ref{fig:si_throughput_44} for the $2\!\times\!2$ and $4\!\times\!4$ cases for LDPC code-rates (1/3, 1/2, 2/3) corresponding to low, medium, and high code-rates, respectively, The proposed AI-ICED receiver is trained at SNR points (10dB, 14dB, 16dB) to cover different SNR range and code-rates. As shown, the throughputs achieved using SI-DMRS (represented by the yellow curves) combined with the AI-ICED receiver consistently outperform the 5G-NR baseline (the green curves), even though both utilize the proposed AI-ICED receivers. Notably, as the SNR increases, the throughput of the 5G-NR approach saturates due to DMRS overheads, whereas transmissions utilizing SI-DMRS achieve full throughput.

Conversely, when using genie CSI, the AI-ICED receiver simplifies to an AI-based MIMO detector, performing (illustrated by the blue curves) nearly as well as the performance bound, namely the genie-CSI-aided MLD (the purple curves). This highlights the effectiveness of AI-ICED in approaching optimal detector performance with genie CSI. Additionally, compared against the baseline genie-CSI-aided LMMSE detector (the red curves), the proposed AI-ICED receiver demonstrates superior performance in middle-to-high SNR regions where detector capability is critical.

\section{Summary}

We have introduced a Transformer encoder based AI-ICED receiver that overcomes DMRS overheads in MIMO-OFDM systems. By integrating the latest VWN, TPA, and MoE architectures, the proposed AI-ICED receiver design reformulates CE and MD as a joint contextual learning process, utilizing a PGFC bridge and an iterative unfolded cascade that is highly resilient to pilot sparsity. With the AI-ICED receiver, both the CE-MSE and BLER converge rapidly within a few iterations, yielding significant performance gains. Furthermore, it achieves detection performance close to that of the optimal MLD when given genie CSI input. Validated by throughput envelopes, the SI-DMRS scheme, combined with the AI-ICED receiver, delivers higher throughput and SE compared to conventional 5G-NR baselines with non-superimposed DMRS and data transmission.

A key question regarding the SI-DMRS scheme involves the power allocation between DMRS and data symbols, as CE and MD are entangled. We have derived the CE-MSE and MD-MSE in approximated closed forms and developed an analytical framework to directly evaluate the final equilibrium states under an ICED process, eliminating the need for numerical simulations. Studies indicate that allocating a power factor of around $0.8$ to DMRS and $0.2$ to data provides an effective design. Furthermore, we have extended the analysis to cases where the power on REs with SI-DMRS can be boosted by borrowing power from remaining REs, yielding substantial additional gains. To evaluate the final decoding performance, we have adopted MIESM to combine MD-MSEs from the two sets of REs: those containing SI-DMRS and those without.

\section*{Appendices}
\subsection{Derivations of CE-MSE}
Letting 
\bea \vec{z}_1\!=\!\sqrt{\frac{(1-b)k}{N_{\text{t}}}}(\vec{H}\vec{x}-\tilde{\vec{H}}\tilde{\vec{x}})+\vec{w},\eea
 and since the estimation error $\vec{\Delta\! H}$ is orthogonal to $\tilde{\vec{H}}$ from the orthogonality principle, it holds that $\mathbb{E}\{\vec{\Delta\! H}\tilde{\vec{H}}\rmh\}\!=\!\vec{0}$. Similarily, $\mathbb{E}\{\vec{\Delta \!x}\tilde{\vec{x}}\rmh\}\!=\!\vec{0}$. Hence, it holds that
\bea 
\mathbb{E}\{\vec{z}_1\vec{z}_1\rmh\} &\!\!\!\!=\!\!\!\!&\frac{(1-b)k}{N_{\text{t}}}\bigg(\mathbb{E}\{\vec{H}\vec{H}\rmh\}\mathbb{E}\{\Delta\!\vec{x}\Delta\!\vec{x}\rmh\}   \notag \\
&& \!\!\!\!+\mathbb{E}\{\Delta\!\vec{H}\Delta\!\vec{H}\rmh\} \mathbb{E}\{\vec{x}\vec{x}\rmh\} \notag \\
&& \!\!\!\! -\mathbb{E}\{\Delta\!\vec{H}\Delta\!\vec{H}\rmh\}\mathbb{E}\{\Delta\!\vec{x}\Delta\!\vec{x}\rmh\} \bigg) \!+\! \sigma^2\vec{I}  \notag \\
&\!\!\!\!=\!\!\!\!&\big( k(1-b)(p+q-pq)+\sigma^2 \big)\vec{I},
\eea
where
\bea 
\mathbb{E}\{\Delta\!\vec{H}\Delta\!\vec{H}\rmh\} \}&\!\!\!\!=\!\!\!\!&N_{\text{t}}\mathbb{E}\{\Delta\!\vec{h}\Delta\!\vec{h}\rmh\} \}=N_{\text{t}}p\vec{I}  \\
\mathbb{E}\{\Delta\!\vec{x}\Delta\!\vec{x}\rmh\}  \} &\!\!\!\!=\!\!\!\!& q\vec{I}.
\eea
With the LMMSE estimator, the CE-MSE equals
\bea \mathbb{E}\{\vec{\Delta\! h}\vec{\Delta\! h}\rmh\} &\!\!\!\!=\!\!\!\!&\!\left(\! bk\rmh \mathbb{E}\{\vec{z}_1\vec{z}_1\rmh\}^{-1} \!+\!\vec{I} \!\right)^{-1} \notag \\
&\!\!\!\!=\!\!\!\!&\left(\! \frac{bk}{ k(1-b)(p+q-pq)+\sigma^2} \!+\!1 \!\right)^{-1}\!\vec{I}\notag \\
&\!\!\!\!=\!\!\!\!& \frac{k(1-b)(p+q-pq)+\sigma^2}{bk+k(1-b)(p+q-pq)+\sigma^2}\vec{I}.
 \eea

\subsection{Derivations of MD-MSE}

Similar, letting 
\bea \vec{z}_2\!=\!\sqrt{\frac{(1-b)k}{N_{\text{t}}}}\vec{\Delta\! H}\vec{x}\!+\!\sqrt{bk}\vec{\Delta\! h}s\!+\!\vec{w},\eea
it holds that
\bea 
\mathbb{E}\{\vec{z}_2\vec{z}_2\rmh\} \!\!&\!\!\!\!=\!\!\!\!&\!\!\frac{(1\!-\!b)k}{N_{\text{t}}}\mathbb{E}\{\Delta\!\vec{H}\Delta\!\vec{H}\rmh\} \mathbb{E}\{\vec{x}\vec{x}\rmh\} \notag \\
&&\!\!+bk\mathbb{E}\{\Delta\!\vec{h}\Delta\!\vec{h}\rmh\} \!+\! \sigma^2\vec{I}  \notag \\
&\!\!\!\!=\!\!\!\!&\!(1\!-\!b)k\mathbb{E}\{\Delta\!\vec{h}\Delta\!\vec{h}\rmh\}+bk\mathbb{E}\{\Delta\!\vec{h}\Delta\!\vec{h}\rmh\} \!+\! \sigma^2\vec{I} \notag \\
&\!\!\!\!=\!\!\!\!& (kp+\sigma^2)\vec{I}.
\eea
With an LMMSE estimator, the MD-MSE equals
\bea \mathbb{E}\{\vec{\Delta \!x}\vec{\Delta \!x}\rmh\} &\!\!\!\!=\!\!\!\!&\!\left(\! \frac{(1-b)k}{N_{\text{t}}}\tilde{\vec{H}}\rmh \mathbb{E}\{\vec{z}_2\vec{z}_2\rmh\}^{-1} \tilde{\vec{H}}\!+\!\vec{I} \!\right)^{-1} \notag \\
&\!\!\!\!=\!\!\!\!&\left(\! \frac{(1-b)k}{N_{\text{t}}(kp+\sigma^2)}\tilde{\vec{H}}\rmh \tilde{\vec{H}}\!+\!\vec{I} \!\right)^{-1} .
\eea

Note that an instantaneous MD-MSE depends on the MIMO channel realization $\vec{H}$, although the expectation has been taken over $\vec{x}$, $\vec{w}$ and $\vec{\Delta h}$. By Jensen's inequality, the ergodic MD-MSE with taking expectation over $\vec{H}$ holds that
\bea 
 \mathbb{E}_{\vec{H}}\{\mathbb{E}\{\vec{\Delta\! x}\vec{\Delta\! x}\rmh\} \}\!\geq\!\left(\! \frac{(1-b)k}{N_{\text{t}}(kp+\sigma^2)} \mathbb{E}\{\tilde{\vec{H}}\rmh \tilde{\vec{H}}\}\!+\!\vec{I} \!\right)^{-1}  .
 \eea
Noting that
 \bea 
\mathbb{E}\{\vec{H}\rmh \vec{H}\}\!=\! \mathbb{E}\{\vec{H}\rmh \vec{H}\}\!-\!\mathbb{E}\{\vec{\Delta\! H}\rmh \vec{\Delta\! H}\}=N_{\text{t}}(1-q)\vec{I},
 \eea
the ergodic MD-MSE is bounded as
 \bea
 \mathbb{E}_{\vec{H}}\{\mathbb{E}\{\vec{\Delta\! x}\vec{\Delta\! x}\rmh\} \}\!\geq\! \frac
{kp+\sigma^2}{k(1-b+bp)+\sigma^2}\vec{I}.
 \eea
In order to analyse the general iterative behaviour, we approximate $\mathbb{E}\{\vec{\Delta x}\vec{\Delta x}\rmh\}$ by its bound, and assuming
\bea 
\mathbb{E}\{\vec{\Delta\!x}\vec{\Delta\! x}\rmh\} \approx  \frac
{kp+\sigma^2}{k(1-b+bp)+\sigma^2}\vec{I}.
\eea

\subsection{Proof of Property~1}

Note that the equilibrium state of CE-MSE and MD-MSE is defined by the equations
\bea 
p&\!\!\!\!=\!\!\!\!&\frac{\gamma N_{\text{t}}}{M}\frac{k(1-b)(p+q-pq)+\sigma^2}{kb+k(1-b)(p+q-pq)+\sigma^2}, \\
q&\!\!\!\!=\!\!\!\!& \frac{kp+\sigma^2}{k(1-b+pb)+\sigma^2}.
\eea
With substitutions $u = 1-b$, $v = \frac{\sigma^{2}}{k}$, and $\rho = \frac{\gamma N_{t}}{M}$, it holds that
\bea
p+q-pq= \frac{p(u+1) - u p^{2} + v}{u + pb + v}.
\eea

Inserting it into the $p$-equation and collecting powers of $p$ yields the cubic polynomial
\bea 
A p^{3} + B p^{2} + C p + D = 0,
\eea
with coefficients $A, B, C, D$ defined in Property~1.

\subsection{Derivations of MD-MSE on REs without DMRS}
Letting 
\bea \vec{z}_3\!=\!\sqrt{\frac{a}{N_{\text{t}}}}\vec{\Delta \!H}\vec{x}\!+\!\vec{w},\eea
it holds that
\bea 
\mathbb{E}\{\vec{z}_3\vec{z}_3\rmh\} \!\!&\!\!\!\!=\!\!\!\!&\!\!\frac{a}{N_{\text{t}}}\mathbb{E}\{\Delta\!\vec{H}\Delta\!\vec{H}\rmh\} \mathbb{E}\{\vec{x}\vec{x}\rmh\} \!+\! \sigma^2\vec{I}  \notag \\
&\!\!\!\!=\!\!\!\!& (ap+\sigma^2)\vec{I}.
\eea
With an LMMSE estimator, the MD-MSE equals
\bea \mathbb{E}\{\vec{\Delta\! x}\vec{\Delta\! x}\rmh\} &\!\!\!\!=\!\!\!\!&\!\left(\! \frac{a}{N_{\text{t}}}\tilde{\vec{H}}\rmh \mathbb{E}\{\vec{z}_3\vec{z}_3\rmh\}^{-1} \tilde{\vec{H}}\!+\!\vec{I} \!\right)^{-1} \notag \\
&\!\!\!\!=\!\!\!\!&\left(\! \frac{a}{N_{\text{t}}(ap+\sigma^2)}\tilde{\vec{H}}\rmh  \tilde{\vec{H}}\!+\!\vec{I} \!\right)^{-1}.
\eea
Following the same argumentations in Appendix-B, we approximate it as 
\bea \mathbb{E}\{\vec{\Delta \!x}\vec{\Delta\! x}\rmh\} \!\approx\! \frac{ap+\sigma^2}{a(1+p)+\sigma^2} .
\eea

\subsection{MIESM}

For a modulation order $M$ (number of bits carried by each constellation symbol), the constellation constrained MI mapping function $f(\theta)$ is commonly approximated as
\bea 
f(\gamma) \approx 
\log_2(M)\left(1 - \sum_{k=1}^{K} a_k e^{-b_k \theta}\right),
\eea
where 
\bea \sum_{k=1}^{K} a_k =1.
\eea

For the widely used exponential fit with $K\!=\!3$, the paramters are listed in Table~\ref{table:mi}.

\subsection{VWN}

Standard Transformer encoders use a fixed hidden dimension ($d_{\mathrm{model}}$) across their layers. Increasing this dimension to handle complex MIMO-OFDM wireless channels causes a quadratic rise in computing cost. As shown in Fig.~\ref{fig:vwn}, VWN separate memory capacity from computational width. The input token $\vec{X}_{\mathrm{token}}^{(i)}$ is expanded to a larger dimension $d_{\mathrm{virtual}} = (n/m) \cdot d_{\mathrm{model}}$ (where $n > m$) and reshaped into $n$ blocks:$$\vec{X}_{\mathrm{block}}^{(i)} \in \mathbb{R}^{B \times L \times n \times d_{\mathrm{b}}}$$Here, the block dimension is $d_{\mathrm{b}} = d_{\mathrm{model}}/m$.

Inside each VWN layer, heavy non-linear operations $\mathcal{F}(\cdot)$ (such as TPA or MoE) are restricted to a narrow $m$-slot computational subspace. Dynamic generalized hyper-connections (DGHC)~\cite{Seed25} manage this interaction. A lightweight linear projection and Tanh-activation on the normalized input dynamically generate a write matrix $\mathcal{B} \!\in\! \mathbb{R}^{m \times n}$ and a joint transformation matrix $\mathcal{A}\! \in\! \mathbb{R}^{(m+n) \times n}$. Horizontally partitioning $\mathcal{A}$ produces the read component $\mathring{\mathcal{A}} \!\in\! \mathbb{R}^{m \times n}$ and the decay residual $\hat{\mathcal{A}}\! \in\! \mathbb{R}^{n \times n}$. The core read-compute-write cycle is executed as
\bea \label{vwm}
    \mathbf{X}_{\text{out}}^{(i)} = \mathcal{B}^\top \mathcal{F}\big(\mathring{\mathcal{A}} \mathbf{X}_{\text{block}}^{(i)}\big) + \hat{\mathcal{A}} \mathbf{X}_{\text{block}}^{(i)}.
\eea
It outlines the data flow: the network tracks complex multi-path residuals across the $n$-dimensional highway, while isolating expensive dense computations to the efficient $m$-dimensional subspace through dynamic gating.

\subsection{TPA}

While the VWN architecture manages macroscopic memory capacity, each layer's core process relies on TPA. Standard Multi-Head Attention (MHA)~\cite{AV23} uses monolithic parameter matrices, limiting its inductive bias. TPA addresses this by factorizing queries, keys, and values into sums of contextual tensor products~\cite{YZ26}, improving parameter efficiency and injecting physical priors.

\textit{Contextual Factorization:}
Given the intermediate state $\mathbf{X}\! \in\! \mathbb{R}^{B \times L \times d_{\mathrm{model}}}$ inside a VWN computational slot, TPA decomposes query, key, and value generation into independent latent factor maps. Let $h$ denote the number of attention heads, $d_{\mathrm{h}} \! =\!  d_{\mathrm{model}}/h$ the per-head dimension, and $r_{\mathrm{q}}, r_{\mathrm{k}}$ the rank constraints for queries and keys/values, respectively. For queries, two linear projections produce head-specific mixing weights $\mathbf{A}_{\mathrm{Q}} \! \in \! \mathbb{R}^{B \times L \times h \times r_{\mathrm{q}}}$ and a shared feature basis $\mathbf{B}_{\mathrm{Q}}\!  \in\!  \mathbb{R}^{B \times L \times r_{\mathrm{q}} \times d_{\mathrm{h}}}$ such that
\bea
    \mathbf{A}_{\mathrm{Q}}(\mathbf{X}) \!\!  &=&\!  \! \mathbf{X} \mathbf{W}_{\mathrm{A}}^{\mathrm{Q}}, \label{TPA_AQ} \\
    \mathbf{B}_{\mathrm{Q}}(\mathbf{X}) \!\!  &=&\!  \!\mathbf{X} \mathbf{W}_{\mathrm{B}}^{\mathrm{Q}}, \label{TPA_BQ}
\eea
where $\mathbf{W}_{\mathrm{A}}^{\mathrm{Q}} \! \in\!  \mathbb{R}^{d_{\mathrm{model}} \times (h \cdot r_{\mathrm{q}})}$ and $\mathbf{W}_{\mathrm{B}}^{\mathrm{Q}}\!  \in\!  \mathbb{R}^{d_{\mathrm{model}} \times (r_{\mathrm{q}} \cdot d_{\mathrm{h}})}$ are factor weight matrices. Analogous projections with rank $r_{\mathrm{k}}$ yield ($\mathbf{A}_{\mathrm{K}}, \mathbf{B}_{\mathrm{K}}$) and ($\mathbf{A}_{\mathrm{V}}, \mathbf{B}_{\mathrm{V}}$).

\textit{Physical Adaptation with 2D-RoPE and Rank Scaling:} Standard 1D rotary position embeddings~\cite{su24} fail to preserve the 2D coherence patterns of MIMO-OFDM channels. To fix this, we treat the sequence axis as a 2D physical grid and apply $\mathrm{RoPE}_{\mathrm{2D}}$ to the feature bases before tensor contraction:
\bea
    \tilde{\mathbf{B}}_{\mathrm{Q}}  \!\!  &=&\!  \! \mathrm{RoPE}_{\mathrm{2D}}\bigl(\mathbf{B}_{\mathrm{Q}}\bigr), \\
    \tilde{\mathbf{B}}_{\mathrm{K}}  \!\!  &=&\!  \! \mathrm{RoPE}_{\mathrm{2D}}\bigl(\mathbf{B}_{\mathrm{K}}\bigr).
\eea
The full query tensor $\mathbf{Q} \! \in\! \mathbb{R}^{B \times L \times h \times d_{\mathrm{h}}}$ is reconstituted via tensor contraction
\begin{equation}
    \mathbf{Q} = \mathbf{A}_{\mathrm{Q}}(\mathbf{X}) \tilde{\mathbf{B}}_{\mathrm{Q}}.
\end{equation}
For the $i$-th head at the $t$-th token, this represents a sum of outer products $\mathbf{Q}_{t}^{(i)} \!=\! \sum_{j=1}^{r_{\mathrm{q}}} \mathbf{a}_{t,j}^{(i)} \otimes \mathbf{b}_{t,j}$. 

\textit{Scaled Dot-Product Attention:}
Using the factorized tensors $\mathbf{Q}, \mathbf{K}, \mathbf{V} \!\in\! \mathbb{R}^{B \times L \times h \times d_{\mathrm{h}}}$, attention is computed per head as
\bea
    \mathrm{head}_i = \mathrm{Softmax}\!\left( \frac{\mathbf{Q}_i \mathbf{K}_i^\top}{\sqrt{d_{\mathrm{h}}}} \right)\! \mathbf{V}_i,
\eea
where $\mathbf{Q}_i, \mathbf{K}_i, \mathbf{V}_i \in \mathbb{R}^{B \times L \times d_{\mathrm{h}}}$ denote the slices along the head dimension. The outputs are concatenated and projected via $\mathbf{W}_{\mathrm{O}} \!\in\! \mathbb{R}^{(h \cdot d_{\mathrm{h}}) \times d_{\mathrm{model}}}$ as
\bea
    \mathrm{TPA}(\mathbf{Q}, \mathbf{K}, \mathbf{V}) = \mathrm{Concat}\bigl(\mathrm{head}_1, \dots, \mathrm{head}_h\bigr) \, \mathbf{W}_{\mathrm{O}}.
\eea
This output completes the TPA layer before routing to the subsequent MoE block.

\subsection{MoE}

After the TPA process, the outputs pass through a MoE feed-forward block adapted from the DeepSeekMoE framework~\cite{ds25}. The expert pool is divided into $K_s$ always-active shared experts and $K_r$ selectively routed experts, both utilizing a SwiGLU-based feed-forward structure~\cite{NS20}. For an input token $\vec{x} \!\in\! \mathbb{R}^{d_{\text{model}}}$, the MoE forward pass is given by
\bea
    \mathbf{y} = \sum_{k=1}^{K_s} \text{MLP}_s^{(k)}(\mathbf{x}) + \sum_{k \in \mathcal{K}} g_k(\mathbf{x}) \, \text{MLP}_r^{(k)}(\mathbf{x}),
\eea
where $\mathcal{K}$ represents the set of top-$K_r$ experts selected by a lightweight gating network, with gating scores computed via a linear projection followed by a Sigmoid activation $\mathbf{s} \!=\! \text{Sigmoid}(\mathbf{x}\mathbf{W}_{\text{route}})$, and the top-$K_r$ entries are normalized to yield the final routing weights:
\bea
    g_k(\mathbf{x}) = \frac{s_k}{\sum_{k \in \mathcal{K}} s_k + \epsilon}, \quad k \in \mathcal{K},
\eea
where $\epsilon$ is a small constant ensuring numerical stability.

\subsection{NN Configuration of the AI-ICED Receiver}

The proposed AI-ICED receiver is configured using the hyperparameters listed in Table~\ref{tab:model_config}. The VWN chassis is configured with $m \!=\! 2$ memory slots and $n \!=\! 3$ active computational slots per layer, producing an expanded virtual dimension $d_{\text{virtual}}\! =\! (n/m) \cdot d_{\text{model}}\!=\!768$. The TPA sub-layer operates with a query rank $r_q \!=\! 16$ and a key/value rank $r_k \!=\! 16$. The MoE sub-layer employs $N_s \!=\! 1$ shared expert, $N_r \!=\!4$ routed experts, and $K_r \!= \!2$ active experts per token. The per-expert hidden dimension scaled by a factor of $(8/3)/(K_r \!+\! N_s)$ to maintain equivalence with a dense SwiGLU FFN~\cite{NS20}. To effectively optimize the deep cascaded Transformer backbone, a hybrid optimization strategy is employed: Multi-dimensional weight matrices are updated using the momentum-based Muon optimizer~\cite{jKJ24}; while one-dimensional parameters, such as layer normalizations and biases, are routed to AdamW~\cite{LH19}. The learning rate follows a linear warmup at first $10\%$ steps, followed by cosine annealing down to a minimum ratio of $0.01$ of the peak value~\cite{PG18, LH17}.

\newpage

\begin{table*}[t]
    \centering
    \small
    \caption{Dominant inference complexity per stage.}
        \vspace{-4mm}
    \label{tab:complexity}
    \renewcommand{\arraystretch}{0.8}
    \begin{tabular}{l p{4cm} p{4cm}}
        \toprule
        \textbf{Operation} & \textbf{Dense Transformer block} & \textbf{Proposed backbone} \\
        \midrule
        Latent memory width 
        & $d_{\mathrm{model}}$
        & $d_{\mathrm{virtual}} = (n/m)d_{\mathrm{model}}$ \\

        Q/K/V projections 
        & $\mathcal{O}(L d_{\mathrm{model}}^{2})$
        & $\mathcal{O}\!\left(L d_{\mathrm{model}}(r_q + 2r)\right)$ \\

        FFN computation 
        & $\mathcal{O}(L d_{\mathrm{model}} d_{\mathrm{ff}})$
        & $\mathcal{O}(L d_{\mathrm{model}} d_{\mathrm{ff}})$ \\

        Attention score 
        & $\mathcal{O}(L^{2} d_{\mathrm{model}})$
        & $\mathcal{O}(L^{2} d_{\mathrm{model}})$ \\
        \bottomrule
    \end{tabular}
        \vspace{-15mm}
\end{table*}

\begin{table*}
    \centering
    \caption{System parameters.}
            \vspace{-4mm}
    \label{tab:sys_params}
    \small
        \renewcommand{\arraystretch}{0.8}
    \begin{tabular}{l l c}
        \hline
        \textbf{Symbol} & \textbf{Description} & \textbf{Value} \\
        \hline
        $N_{\mathrm{r}} \times N_{\mathrm{t}}$ & MIMO size & $2 \times 2$, $4 \times 4$ \\
        $\mathcal{A}$ & Modulation scheme & 16QAM, 64QAM \\
        $N_{\mathrm{sym}}$ & Number of OFDM symbols carrying data & $12$ \\
        $N_{\mathrm{sc}}$ & Active subcarriers & $24$ \\
         $N$ & Number of REs carrying only data & $240$ \\
        $M$ & Number of REs carrying both SI-DMRS and data & $48$ \\
         $L$ & Token lenght for the AI-ICED receiver & $288$ \\
        Codes & LDPC & Code-rates $1/3$, $1/2$, $2/3$ \\
        $(a, b)$ & Power allocation factors & $(a=1, b=0.8)$ \\
        Channels & 3GPP channel models & ETU-70Hz, EPA-5Hz \\
        \hline
    \end{tabular}
    \vspace{-15mm}
\end{table*}

\begin{table*} 
\centering
\small
\caption{Three-term exponential MI mapping parameters.}
        \vspace{-4mm}
        \label{table:mi}
\begin{tabular}{c|cccccc}
\textbf{Modulation} & $a_1$ & $b_1$ & $a_2$ & $b_2$ & $a_3$ & $b_3$ \\ \hline
QPSK   & 0.9810 & 1.0420 & 0.0190 & 4.2500 & 0.0000 & 0.0000 \\
16-QAM & 0.6053 & 0.2845 & 0.3541 & 1.2562 & 0.0406 & 5.3784 \\
64-QAM & 0.4076 & 0.0717 & 0.3951 & 0.3556 & 0.1973 & 1.8315 \\
\end{tabular}
\vspace{30mm}
\end{table*}

\begin{table*}
\caption{Hyperparameter Configuration of the Proposed AI-ICED Receiver}
        \vspace{-4mm}
\label{tab:model_config}
\centering
\small
\renewcommand{\arraystretch}{0.8}
\begin{tabular}{l l c}
    \toprule
    \textbf{Symbol} & \textbf{Description} & \textbf{Value} \\
    \midrule
    \multicolumn{3}{c}{\textit{Transformer Backbone}} \\
    $d_{\text{model}}$ & Latent dimension & $512$ \\
    $N_{\text{head}}$ & Number of attention heads & $8$ \\
    $m/n$ & VWN memory / compute slots & $2/3$ \\
    $d_{\text{virtual}}$ & Virtual memory width $(n/m) \cdot d_{\text{model}}$ & $768$ \\
    $r_q/r$ & TPA query rank / key-value rank & $16/16$ \\
    $N_s/N_r$ & Shared / routed experts (MoE) & $1/4$ \\
    $K_r$ & Active routed experts per token & $2$ \\
    Expert factor & Per-expert hidden factor $(8/3)/(K_r + N_s)$ & $8/9$ \\
    Dropout & Dropout probability & $0.1$ \\
    \midrule
    \multicolumn{3}{c}{\textit{Cascade Architecture}} \\
    $N_{\text{CE}}^{(0)} / N_{\text{MD}}^{(0)}$ & Bootstrap CE / MD encoder layers & $3/3$ \\
    $N_{\text{CE}}^{(i)} / N_{\text{MD}}^{(i)}$ & Refinement CE / MD encoder layers & $6/6$ \\
    $N_{\text{it}}$ & Number of ICED iteration & $1$ \\
    \midrule
    \multicolumn{3}{c}{\textit{Training Protocol}} \\
    Epochs & Total training epochs & $100$ \\
    Batch size & Samples per batch & $16$ \\
    Optimizer & Hybrid & Muon + AdamW \\
    LR (Muon) & Peak learning-rate for multi-dimensional weights & $2 \!\times\! 10^{-3}$ \\
    LR (AdamW) & Peak learning-rate for embeddings, norms, biases & $3 \!\times \!10^{-4}$ \\
    Weight decay & $\ell_2$ regularization penalty & $0.01$ \\
    $\beta_1$ &     Loss balance between bit cross-entropy and CE-MSE & $0.1$ \\
    $\beta_2$ & Auxiliary stage weight & $0.3$ \\
    \bottomrule
\end{tabular}
\vspace{0mm}
\end{table*}

%


\begin{thebibliography}{99}

\bibitem{fu26}
Z.~Fu, ``Deep in-context learning (ICL) for wireless communications,'' Master's thesis, Dept. of electrical and information technology, Lund University, Lund, Sweden, Jun. 2026.

\bibitem{Wang2026AI}
X. Wang, L. Lu, Q. Li, Q. Sun, N. Shi, Z. Chen, and T. Sun, ``A task-driven design approach for 6G AI-native architecture,'' \textit{Engineering}, vol. 56, no. 1, pp. 87--103, 2026.

\bibitem{Wu2024}
Y. Wu \textit{et al.}, ``A comprehensive review of AI-native 6G: Integrating semantic communications, reconfigurable intelligent surfaces, and edge intelligence for next-generation connectivity,'' \textit{Front. Commun. Netw.}, vol. 5, p. 1655410, 2024.

\bibitem{QJ19}
Z. Qin, H. Ye, G. Y. Li, and B.-H. F. Juang, ``Deep learning in physical layer communications,'' \emph{IEEE Wireless Commun.}, vol. 26, no. 2, pp. 93--99, Apr. 2019.

\bibitem{Sajid2026}
J. Sajid \textit{et al.}, ``Empowering embodied AI in 6G networks: Architecture, enablers, and open challenges,'' \textit{arXiv preprint arXiv:2606.20592}, 2026.

\bibitem{OH17}
T. O’Shea and J. Hoydis, ``An introduction to deep learning for the physical layer,'' \emph{IEEE Trans. Cogn. Commun. Netw.}, vol. 3, no. 4, pp. 563--575, Dec. 2017.

\bibitem{lin2023}
X. Lin, ``An overview of the 3GPP study on artificial intelligence for 5G new radio,'' \textit{arXiv preprint arXiv:2308.05315}, 2023.

\bibitem{QS24}
X. Qin, S. Hu, J. Zhang, J. Qian, and H. Wang, ``AI receiver design with deep learning based channel estimation and MIMO detection,'' in \emph{Proc. IEEE PIMRC}, 2024, pp. 1--7.

\bibitem{Li2025}
X. Li, X. Zhou, Y. Cao, J. Zhang, C.-K. Wen, X. Li, and S. Jin, ``Learning-aided iterative receiver for superimposed pilots: Design and experimental evaluation,'' \textit{arXiv preprint arXiv:2507.10074}, 2025.

\bibitem{QS26}
X. Qin and S. Hu, ``Dual-attention based 3D channel estimation,'' \textit{arXiv preprint arXiv:2604.01769}, 2026.

\bibitem{C23}
S. Cammerer \textit{et al.}, ``A neural receiver for 5G NR multi-user MIMO,'' in \emph{Proc. IEEE Globecom Workshops (GC Wkshps)}, Kuala Lumpur, Malaysia, 2023, pp. 329--334.

\bibitem{Hu26}
S. Hu, ``Invariant transformation and resampling based epistemic-uncertainty reduction,'' \textit{arXiv preprint arXiv:2602.23315}, 2026.

\bibitem{BB26}
F. B. Saghezchi, M. Pourghasemian, B. Ding, A. Abdi, B. Lee, and A. Baron, ``AI-native radio transceiver signal processing for next-generation mobile communication systems,'' \textit{IEICE Trans. Commun.}, vol. E109-B, no. 4, pp. 555--572, Apr. 2026.

\bibitem{HL18}
H. He, C.-K. Wen, S. Jin, and G. Y. Li, ``A model-driven deep learning network for MIMO detection,'' in \emph{Proc. IEEE Glob. Conf. Signal Inf. Process. (GlobalSIP)}, Anaheim, CA, USA, 2018, pp. 584--588.

\bibitem{MA24}
A. Mazumdar, C. N. Manchon, O. E. Barbu, and R. O. Adeogun, ``Comparative evaluation of model based deep learning receivers in coded MIMO systems,'' in \emph{Proc. IEEE Veh. Technol. Conf. (VTC-Fall)}, 2024, pp. 1--7.


\bibitem{HW98}
M.~-H.~Hsieh and C.~-H.~Wei, ``Channel estimation for {OFDM} systems based on comb-type pilot arrangement in frequency selective fading channels,'' \emph{IEEE Trans. Consum. Electron.}, vol. 44, no. 1, pp. 217--225, Feb. 1998.

\bibitem{DD04}
D. Wubben, R. Bohnke, V. Kuhn, and K.-D. Kammeyer, ``MMSE-based lattice-reduction for near-ML detection of MIMO systems,'' in \emph{Proc. ITG Workshop Smart Antennas}, 2004, pp. 106--113.

\bibitem{HR17}
S.~Hu and F.~Rusek, ``A soft-output {MIMO} detector with achievable information rate based partial marginalization,'' \emph{IEEE Trans. Signal Process.}, vol. 65, no. 6, pp. 1622--1637, Mar. 2017.

\bibitem{UV17}
K. Upadhya, S. A. Vorobyov, and M. Vehkap{\"e}{\"a}, ``Superimposed pilots are superior for mitigating pilot contamination in massive MIMO,'' \textit{IEEE Trans. Signal Process.}, vol. 65, no. 11, pp. 2917--2932, Jun. 2017.

\bibitem{SB25}
S. Rezaie, M. Honkala, D. Korpi, D. C. Melgarejo, T. Izydorczyk, D. Gold, and O.-E. Barbu, ``Superimposed DMRS for spectrally efficient 6G uplink multi-user OFDM: Classical vs AI/ML receivers,'' \textit{arXiv preprint arXiv:2506.20248}, 2025.

\bibitem{LJ25}
X. Li, X. Zhou, J. Zhang, C.-K. Wen, and S. Jin, ``AI-driven iterative receiver for superimposed pilot Schemes in MIMO-OFDM systems,'' in \emph{Proc. IEEE WCNC}, 2025, pp. 1--6.

\bibitem{ying26}
K.~Ying \emph{et~al.}, ``Conditional diffusion model-driven sassive {MIMO} iterative detection,'' in \emph{Proc. IEEE Int. Conf. Commun. (ICC)}, Glasgow, United Kingdom, 2026, pp. 1--6.

\bibitem{AH08}
M.~Abuthinien, S.~Chen, and L.~Hanzo, ``Semi-blind joint maximum likelihood channel estimation and data detection for {MIMO} systems,'' \emph{IEEE Signal Process. Lett.}, vol. 15, pp. 202--205, 2008.

\bibitem{ds25}
DeepSeek-AI \textit{et al.}, ``DeepSeek-V3.2: pushing the frontier of open large language models,'' \textit{arXiv preprint arXiv:2512.02556}, 2025.

\bibitem{Qwen25}
A. Yang \textit{et al.}, ``Qwen3 technical report,'' \textit{arXiv preprint arXiv:2505.09388}, 2025.

\bibitem{kim2.5}
Kimi Team \textit{et al.}, ``Kimi K2.5: Visual Agentic Intelligence,'' \textit{arXiv preprint arXiv:2602.02276}, 2026.

\bibitem{SS25}
Z. Song, M. Zecchin, B. Rajendran, and O. Simeone, ``Turbo-ICL: In-context learning-based turbo equalization,'' \textit{arXiv preprint arXiv:2505.06175}, 2025.

\bibitem{SR26}
M. Shanmugam, S. Balaraman, and K. Ranganathan, ``Improving channel equalization in cell-free MIMO networks using reconfigurable intelligent surfaces and in-context learning,'' \emph{Ann. Telecommun.}, vol. 81, pp. 259--273, 2026.

\bibitem{ZS24}
M. Zecchin, K. Yu, and O. Simeone, ``In-context learning for MIMO equalization using Transformer-based sequence models,'' in \emph{Proc. IEEE Int. Conf. Commun. Workshops (ICC Workshops)}, 2024, pp. 1573--1578,

\bibitem{Seed25}
Seed \emph{et~al.}, ``Virtual Width Networks,'' \emph{arXiv preprint arXiv:2511.11238}, 2025.

\bibitem{YZ26}
Y. Zhang \textit{et al.}, ``Tensor product attention is all you need,'' \textit{arXiv preprint arXiv:2501.06425}, 2026.

\bibitem{ds24}
D. Dai \textit{et al.}, ``DeepSeekMoE: Towards ultimate expert specialization in mixture-of-experts language models,'' \textit{arXiv preprint arXiv:2401.06066}, 2024.

\bibitem{3gpp38211}
{3GPP}, ``5G; {NR}; Physical channels and modulation,'' {3GPP} Technical Specification TS 38.211, V18.6.0, Apr. 2025.

\bibitem{3gpp36101}
{3GPP}, "Evolved Universal Terrestrial Radio Access ({E-UTRA}); User Equipment ({UE}) radio transmission and reception," {3GPP} Technical Specification TS 36.101, V18.9.0, Apr. 2025.

\bibitem{AV23}
A.~Vaswani \emph{et~al.}, ``Attention is all you need,'' in \emph{Advances Neural Inf. Process. Syst. (NeurIPS)}, vol. 30, 2017, pp. 5998--6008.

\bibitem{su24}
J. Su \textit{et al.}, ``Roformer: Enhanced transformer with rotary position embedding,'' \emph{Neurocomputing}, vol. 568, p. 127063, 2024.

\bibitem{NS20}
N. Shazeer, ``GLU variants improve Transformer,'' \textit{arXiv preprint arXiv:2002.05202}, 2020.

\bibitem{jKJ24}
K. Jordan \textit{et al.}, ``Muon: An optimizer for hidden layers in neural networks,'' 2024. [Online] Available: \textit{https://kellerjordan.github.io/posts/muon/}.

\bibitem{LH19}
I. Loshchilov and F. Hutter, ``Decoupled weight decay regularization,'' \textit{arXiv preprint arXiv:1711.05101}, 2019.

\bibitem{PG18}
P. Goyal \textit{et al.}, ``Accurate, large Minibatch SGD: Training ImageNet in 1 Hour,'' \textit{arXiv preprint arXiv:1706.02677}, 2017.

\bibitem{LH17}
I. Loshchilov and F. Hutter, ``SGDR: Stochastic gradient descent with warm restarts,'' \textit{arXiv preprint arXiv:1608.03983}, 2017.

\bibitem{YZ14}
J.~Yang, H.~Zhao, W.~Wang, and C.~Zhang, ``An effective SINR mapping models for 256QAM in LTE-Advanced system,'' in \emph{IEEE Annu. Int. Symp. Pers., Indoor, Mobile Radio Commun. (PIMRC)}, Washington, DC, USA, 2014, pp. 343--347.

\bibitem{KP08}
N. Kim, Y. Lee, and H. Park, ``Performance analysis of MIMO system with linear MMSE receiver,'' \textit{IEEE Trans. Wireless Commun.}, vol. 7, no. 11, pp. 4474--4478, Nov. 2008.

\end{thebibliography}
\end{document}